\documentstyle[12pt]{article}    
\def\beq{\begin{equation}}
\def\eeq{\end{equation}}
\def\bea{\begin{eqnarray}}
\def\eea{\end{eqnarray}}
\def\bq{\begin{quote}}
\def\eq{\end{quote}}

\def\bq{\begin{quote}}
\def\eq{\end{quote}}

\begin{document}

\baselineskip 24pt

\newcommand{\sheptitle}
{Soft Masses in Strong Unification}

\newcommand{\shepauthor}
{S. F. King$^{\dagger}$
\footnote{On leave of absence from 
Department of Physics and Astronomy,
University of Southampton, Southampton, SO17 1BJ, U.K.}
and G. G. Ross$^\ast $}

\newcommand{\shepaddress}
{$^\dagger$ Theory Division, CERN, CH-1211 Geneva 23, Switzerland\\
$^\ast$ Department of Physics. Theoretical Physics, University of Oxford\\
1 Keble Road, Oxford OX1 3NP, U.K.}

\newcommand{\shepabstract}
{We examine the infra-red structure of 
soft supersymmetry breaking masses of the squarks and sleptons
in supersymmetric (SUSY) strong unification schemes
where $\alpha_{GUT}\sim 0.2-1$.
We show that combinations of soft masses approach
fixed points which leads to very simple predictions for
soft masses at the intermediate scale.
The assumption that the high energy gaugino mass $M_{1/2}$
dominates over the soft scalar masses leads to a strong suppression of
flavour changing neutral currents, and a
low energy SUSY spectrum which is simply predicted in terms of two parameters,
namely $\alpha_{GUT}$ and $M_{1/2}$. Due to the quickly falling gauge
couplings beneath the high energy scale, the low energy spectrum has a
characteristic ``scalar dominated'' signature quite
unlike the standard gaugino dominated MSSM where the right-handed
sleptons are predicted to be rather light.
We also examine the new sources of flavour changing expected in such models, 
and in particular show that the flavour violation coming from the
D-term of a $U(1)_X$ gauged family
symmetry may be reduced to an acceptable level.}

\begin{titlepage}
\begin{flushright}
CERN-TH/98-27\\
OUTP-98-22-P\\
hep-ph/9803463\\
\end{flushright}
\begin{center}
{\large{\bf \sheptitle}}
\bigskip \\ \shepauthor \\ \mbox{} \\ {\it \shepaddress} \\ 
{\bf Abstract} \bigskip \end{center} \setcounter{page}{0}
\shepabstract
\begin{flushleft}
CERN-TH/98-27\\
\today
\end{flushleft}
\end{titlepage}

\section{Introduction}

Models which extend the MSSM through the 
addition of massive multiplets which are 
vectorlike with respect to the Standard Model 
gauge group occur very often in the breakdown 
of Grand Unified theories or in compactified 
string theories. Provided these states fill out 
complete $SU(5)$ representations they lead to an increased gauge 
coupling at the unification scale without disturbing 
the success of the one-loop unification predictions. 
As a result the ratio of the Yukawa couplings to 
gauge couplings are driven quickly to 
infra-red fixed points, offering the possibility 
of a dynamical understanding of the 
pattern of fermion masses. 
It may also happen that the fixed point 
structure corresponds to family independent 
Yukawa couplings offering the 
possibility of a dynamical understanding 
of the flavour problem in supersymmetric theories. 

Motivated by such considerations
unification predictions have been re-examined in theories in which
there is extra matter in complete $SU(5)$ representations at an
intermediate mass scale $M_{I}$ below the unification scale 
\cite{strong1, kolda}.
The extra matter consists of $n_{5}$ copies of $(5+\bar{5})$ plus $n_{10}$
copies of $(10+\bar{10})$ representations which serve to increase the beta
functions above the scale $M_{I}$, resulting in an increased value of the
unified gauge coupling $\alpha _{GUT}$. On the other hand the
unification scale $M_{GUT}\sim 2\times 10^{16}$ GeV is virtually unchanged
from its MSSM value due to an accurate cancellation between the two-loop and
threshold effects \cite{strong1}. The presence of such additional matter is
typical of a certain class of string model in which gauge symmetries are
broken by Wilson lines \cite{strong2}. Moreover such extra matter is welcome
since it may serve to increase the unified coupling to a value which is
large enough to solve the ``dilaton runaway problem'', providing also that
the string scale $M_{string}$ is reduced down to $M_{GUT}$ \`{a} la Witten 
\cite{W, dilaton}.

In such theories the values of the gauge couplings near the unification
scale may be raised into the strong coupling region \cite{strong3},
effectively placing the question of the unification of the gauge couplings
outside of perturbation theory. At first sight this would seem to imply that
all the predictive power of unification is lost. However, as shown in \cite
{strong3}, low energy predictivity is maintained since the steeply falling
gauge couplings are quickly driven to precise fixed point ratios: 
\begin{equation}
\frac{\alpha _{1}}{\alpha _{3}}\rightarrow r_{1}\equiv \frac{b_{3}}{b_{1}},\ \
\frac{\alpha _{2}}{\alpha _{3}}\rightarrow r_{2}\equiv \frac{b_{3}}{b_{2}},  
\label{r}
\end{equation}
where the beta functions are 
\begin{equation}
b_{a}=\left( 
\begin{array}{c}
33/5+n \\ 
1+n \\ 
-3+n
\end{array}
\right)  \label{b}
\end{equation}
where $n=(n_{5}+3n_{10})$. Thus one may take the ratios $r_{1},r_{2}$ as a
boundary condition at the scale $M_{I}$, and hence use them to determine the
low energy measured couplings. In this approach the scale $M_{I}$ is
regarded as an input parameter which may, for a given value of $n$, be fixed
by two of the gauge couplings (say $\alpha _{1}$ and $\alpha _{2}$):
\begin{equation}
\frac{1}{2\pi }\ln \left[ \frac{M_I}{M_Z}\right]\approx
\frac{29.3n-136.9}{5.6n}
\label{MI}
\end{equation}
The third gauge coupling may be predicted at low energies as in the standard
unification picture, and indeed leads to values of $\alpha _{3}(M_{Z})$ in
good agreement with experiment \cite{strong3}. This prediction, which
follows without a conventional scale $M_{GUT}$, originates from the precise
boundary conditions in Eq.\ref{r} at $M_{I}$. The gauge couplings become
non-perturbative at a scale $M_{NP}$, close to the
conventional GUT scale \cite{strong3}. Note that $M_{I}$ is the mass scale
in the superpotential, {\it not} the physical mass of the heavy states which
receive large radiative corrections. Such radiative splitting effects
decouple from the evolution equations for the couplings\cite{shifman}.

The key to the predictive power of this scheme is the steeply falling gauge
couplings in the region $M_{NP}-M_{I}$, which drives the gauge couplings to
their fixed point values at $M_{I}$ in Eq.\ref{r} \cite{strong3}. Similarly
any dimensionless Yukawa couplings which are initially of the same general
order as the gauge couplings will evolve to precise fixed point ratios in
the infra-red similar to the fixed point for the top quark Yukawa coupling 
\cite{PR}. In the MSSM the rate of approach to the
fixed point is not very efficient
since the ratio of the top Yukawa
coupling $h$ to the gauge coupling $g_{3}$ at low energy in the 
one loop is approximation is given by
(assuming negligible bottom Yukawa coupling) 
\begin{equation}
R(M_{Z})=\frac{R^{\ast }}{1+\Delta \left( \frac{R^{*}}{R(M_{GUT})}-1\right) }
\end{equation}
where 
$$
R(M_{Z})\equiv \frac{h^{2}(M_{Z})}{g^2_{3}(M_{Z})}
$$
and
$R^{\ast }=7/18$ is the fixed point ratio in the MSSM and 
$$
\Delta =\left( 
\frac{\alpha _{3}(M_{Z})}{\alpha _{GUT}} \right) ^{(1+16/3b_{3})},
$$
where in
the MSSM the value of $\Delta $ is not that small, $\Delta \approx 1/3$, so
that the rate of approach to the PR fixed point is not that large. 
\footnote{%
In the MSSM if $R^\ast /R(M_{GUT})\ll 1$ then clearly 
$R(M_Z)$ $\approx$ $R^\ast /(1-\Delta )$
which is the so-called quasi-fixed point of Hill \cite{H}.}
However in strong unification, the value of $\Delta $ appropriate to the
region between $M_{NP}$ and $M_{I}$, is typically much smaller due to the
steeply falling couplings, thereby significantly increasing 
the rate of approach to the fixed point. \footnote{%
The fact that small $\Delta $ increases the rate of approach to the PR
fixed point was emphasised by Lanzagorta and Ross \cite{LR}.}

It was shown that if all the third family Yukawa couplings are assumed to be
of order the gauge couplings at high energies, then they efficiently approach
the fixed point and then
one obtains precise predictions for third family Yukawa couplings at the
scale $M_{I}$, and hence precise low energy predictions for third family
masses and the ratio of Higgs vacuum expectation values $\tan \beta $ as a
function of the parameter $n$. In this case $\tan \beta
\approx 46-47$ and the top quark mass exceeds 200 GeV in all cases. 
However an acceptable top mass may be achieved for $%
\tan \beta \approx 1.01-1.3$. As pointed out \cite{strong3},
these predictions for the third family Yukawa couplings are sensitive to
other Yukawa couplings
and, in a particular theory of fermion masses%
\cite{GG}, the presence of large Yukawa couplings involving the first and
second families will affect the low energy predictions of the third family
spectrum and reduce the top mass prediction to acceptable values \cite
{strong3}.

In this paper we shall examine the question of the infra-red behaviour of soft
supersymmetry (SUSY) breaking parameters\cite{LR2} in the framework of
strong unification, focussing in particular on the above mentioned theory of
fermion masses \cite{GG} as a concrete example. 
The purpose of the present
paper is to extend the above 
MSSM analysis to include all three families in the
particularly promising framework of strong unification combined with the
model of fermion masses mentioned above. 
Since the gauge couplings and Yukawa couplings approach fixed points
one might expect that the soft mass parameters will also
approach fixed points in the infrared, leading to enhanced
predictivity of the spectrum at low energies.
As in the MSSM, however, one finds there are combinations of soft masses
which are not suppressed by anomalous dimensions, so the soft masses
at low energies will inevitably depend on physics at high energies.
As a result the fixed point structure, by itself, 
does not give a dynamical explanation of the flavour problem.
On the other hand there are combinations of soft masses which are
heavily suppressed by renormalisation group running, 
and this effect serves drastically to simplify the 
relation between the soft masses at the
intermediate scale and those at the string scale.

A particularly attractive hypothesis that does 
lead to a solution to the flavour problem is that the gaugino mass at the 
string scale energy dominates over the scalar masses at the string scale
(gaugino dominance.) In this case the efficient anomalous dimension
suppression allows the scalar masses at the intermediate scale
to be predicted very simply in terms of the high energy gaugino mass.
The soft scalar mass differences between families, which is the source
of flavour changing neutral currents in SUSY theories, is then
accurately set to zero.
Furthermore, the fixed point structure accurately preserves the 
family independence of the soft scalar masses for all three
families, down to the intermediate scale. 
We discuss the
detailed low energy SUSY spectrum in a particular model of this kind.
Whereas the assumption of gaugino dominance in the minimal supersymmetric
standard model (MSSM) naturally leads to a low energy spectrum
involving heavy gauginos and light squarks and sleptons, in the present model
gaugino dominance leads to a low energy spectrum involving 
light gauginos and heavy squarks and sleptons -- exactly the reverse of the
MSSM spectrum in this case! The reason for this is due to the rapidly
falling gauge couplings between the string scale and the intermediate
scale which causes the gaugino masses to similarly fall.
The scale of the soft scalar masses at the intermediate scale
is set by the gaugino mass at the string scale which is much larger,
and this gives rise to the inverted spectrum which may make
gauge dominance more viable in strong unification than in the
MSSM. For example the right-handed selectron, which is always very light
in the MSSM gauge dominated scenario, is comfortably heavy in the present
model.

The layout of the paper is as follows. In section 2 we introduce the
model, and in section 3 we present the renormalisation group equations
(RGEs) that will form the focus of the rest of the paper.
In section 4 we analyse the fixed point solution these equations,
when summed over flavours. Section 5 discusses the extra subtleties
arising from the additional Higgs couplings that are present in the
model, and describes a simple augmented model with extra Higgs couplings.
In section 6 we analyse the fixed point solutions to the
equations for the differences between the soft scalar masses in flavour space.
In section 7 we discuss the character of the
low energy SUSY spectrum in some detail for a particular example
based on gaugino dominance and large ratio of Higgs 
vacuum expecation values $\tan \beta$.
In section 8 we discuss a further source of flavour violation
coming from the D-term of the gauged family symmetry.
We reserve section 9 for our summary and conclusions.

\section{A Family symmetry}

The theory of fermion masses on which our results are based \cite{GG} relies
on a gauged $U(1)_{X}$ family symmetry \cite{IR}. The $U(1)_{X}$ has a
Green-Schwarz anomaly \cite{GS} and is assumed to be broken close to the
string scale by the vacuum expectation values (VEVs) of standard model
singlet fields including
$\theta $ and $\bar{\theta}$ with $U(1)_{X}$ charges 1 and -1
respectively \cite{IR}. In order to achieve a realistic pattern of masses
the following simple $X$ charges were assumed for the three families of
quarks and leptons (assigning equal charges to each multiplet member $%
Q_{i},U_{i}^{c},D_{i}^{c},L_{i}^{c},E_{i}^{c}$ of the ith family): 
\begin{equation}
\begin{array}{cc}
1^{st}\ \mbox{family:} & X=-4 \\ 
2^{nd}\ \mbox{family:} & X=+1 \\ 
3^{rd}\ \mbox{family:} & X=0
\end{array}
.
\end{equation}
In this model the light fermion masses are generated because the light Higgs
fields of the Standard Model are mixtures of several different Higgs fields
carrying different X charges. In the model we are considering these
additional heavy Higgs fields must belong to the additional vector-like
matter multiplets transforming as complete $SU(5)$ representations. Given
that these new states must have renormalisable Yukawa couplings to the
quarks and leptons they must have $X$ charges to cancel the charge matrix
associated with the matrix of $qq^{c}$ quark charges,
\[
\left( 
\begin{array}{ccc}
-8 & -3 & -4 \\ 
-3 & +2 & +1 \\ 
-4 & +1 & 0
\end{array}
\right) .
\]
To generate these entries we need 9 
(6 if the Higgs couplings are symmetric
\footnote{Note that the matrix of charges is symmetric in this
example, so that in principle a Higgs coupling in a particluar
off-diagonal entry could also couple in the symmetric position.
We shall assume that this does not happen, namely that a particular
Higgs only couples in a particular position. If necessary we can enforce
this by suitable discrete symmetries.}
)
Higgs doublets $H_{U_{ij}}$ coupling to the up sector and 9 (6) Higgs
doublets $H_{D_{ij}}$ coupling to the down and lepton sector with $%
H_{U_{11}} $ having charge $X=8$, and so on. The Higgs $H_{U_{33}}$, $%
H_{D_{33}}$ both have $X=0$ and give the dominant component of the two Higgs
doublets $H_{U}$, $H_{D}$ of the minimal SUSY model. The Higgs doublets with
non-zero $X$ charge are assumed to have masses of order $M_{I}$ due to
couplings to their vector partners. However the 9 Higgs of each type mix via
Frogatt-Nielsen \cite{FN} diagrams involving insertions of the $\theta $ and 
$\bar{\theta}$ fields along the Higgs line, so that at low energies the up
mass matrix for example results from the matrix of operators: 
\begin{eqnarray}
Q_{1}U_{1}^{c}H_{U}({\theta }/M_{I})^{8},\ &Q_{1}U_{2}^{c}H_{U}({\theta }%
/M_{I})^{3},\ &Q_{1}U_{3}^{c}H_{U}({\theta }/M_{I})^{4},  \nonumber \\
Q_{2}U_{1}^{c}H_{U}({\theta }/M_{I})^{3},\ &Q_{2}U_{2}^{c}H_{U}(\bar{\theta}%
/M_{I})^{2},\ &Q_{2}U_{3}^{c}H_{U}(\bar{\theta}/M_{I}),  \nonumber \\
Q_{3}U_{1}^{c}H_{U}({\theta }/M_{I})^{4},\ &Q_{3}U_{2}^{c}H_{U}(\bar{\theta}%
/M_{I}),\ &Q_{3}U_{3}^{c}H_{U}  \label{nonrenops}
\end{eqnarray}
Once the $\theta $ and $\bar{\theta}$ fields acquire VEVs these terms
generate a matrix of Yukawa couplings 
\begin{equation}
\left( 
\begin{array}{ccc}
\epsilon ^{8} & \epsilon ^{3} & \epsilon ^{4} \\ 
\epsilon ^{3} & \epsilon ^{2} & \epsilon \\ 
\epsilon ^{4} & \epsilon & 1
\end{array}
\right) .  \label{epsilonmatrix}
\end{equation}
which makes contact with the textures of the 
mass matrix responsible for some of the 
pattern of light quark masses and 
mixing angles \cite{RRR}. 
Here texture zeroes are due to the appearance of high powers of
the expansion parameter $\epsilon =<\theta >/M_{I}=<\bar{\theta}>/M_{I}$,
where $\epsilon \approx 0.2$.

In the region below the scale $M_{NP}$,
the scale at which the gauge couplings become large, 
but above $M_{I}$ the renormalisable superpotential
contains Yukawa couplings involving the 18 Higgs doublets $H_{U_{ij}}$,$%
H_{D_{ij}}$ which generate the terms of eq(\ref{nonrenops}). These are 
\begin{equation}
W=%
\sum_{i,j=1}^{3}(h_{ij}Q_{i}U_{j}^{c}H_{U_{ij}}
+k_{ij}Q_{i}D_{j}^{c}H_{D_{ij}}+l_{ij}L_{i}E_{j}^{c}H_{D_{ij}})
\label{W}
\end{equation}
There are also renormalisable terms involving singlets which we shall
discuss later.

\section{The renormalisation group equations for soft masses}

The soft SUSY breaking potential has the general form 
\begin{eqnarray}
V_{soft}
&=&-%
\sum_{i,j=1}^{3}(A_{h_{ij}}h_{ij}Q_{i}U_{j}^{c}H_{U_{ij}}
+A_{k_{ij}}k_{ij}Q_{i}U_{j}^{c}H_{D_{ij}}
+A_{l_{ij}}l_{ij}L_{i}E_{j}^{c}H_{D_{ij}}+H.c)
\nonumber \\
&+&\frac{1}{2}\sum_{i,j=1}^{3}m_{H_{U_{ij}}}|H_{U_{ij}}|^{2}+\frac{1}{2}%
\sum_{i,j=1}^{3}m_{H_{D_{ij}}}|H_{D_{ij}}|^{2}  \nonumber \\
&+&\frac{1}{2}%
\sum_{j=1}^{3}(m_{Q_{j}}^{2}|Q_{j}|^{2}+m_{U_{j}^{c}}^{2}|U_{j}^{c}|^{2}
+m_{D_{j}^{c}}^{2}|D_{j}^{c}|^{2}+m_{L_{j}}^{2}|L_{j}|^{2}
+m_{E_{j}^{c}}^{2}|E_{j}^{c}|^{2})
\nonumber \\
&+&\frac{1}{2}\sum_{a=1}^{3}M_{a}\lambda _{a}^{2}+H.c.  \label{Vsoft}
\end{eqnarray}
Note that the $X$ symmetry (assumed to be unbroken down to $M_{I}$) forbids
off-diagonal squark and slepton masses at renormalisable order, moreover
non-renormalisable operators will not enter the RGEs.

The renormalisation group equations (RGEs) for the gauge couplings and
gaugino masses are: 
\begin{equation}
\frac{d\tilde{\alpha}_{a}}{dt}=-b_{a}\tilde{\alpha}_{a}^{2},\ \ \frac{dM_{a}%
}{dt}=-b_{a}M_{a}\tilde{\alpha}_{a}
\end{equation}
where we have defined 
$\tilde{\alpha}_{a}\equiv \frac{g_{a}^{2}}{16\pi ^{2}}$, 
$t\equiv \ln (M_{NP}^2/\mu ^{2})$ 
with $\mu $ being the $\bar{MS}$ scale and $b_{a}$
the beta functions given in Eq.\ref{b}. 
We are interested in solving the
RGEs between the scale $M_{NP}\approx M_{GUT}$
and the intermediate scale $M_I$.

The RGEs for the gauge couplings and gaugino masses have the 
well known solution:
\begin{eqnarray}
\tilde{\alpha}_{a}(t)
& = & \frac{\tilde{\alpha}_{a}(0)}{1+b_a\tilde{\alpha}_{a}(0)t} \nonumber \\
M_{a}(t)
& = & \frac{M_{a}(0)}{1+b_a\tilde{\alpha}_{a}(0)t} 
\label{wellknown}
\end{eqnarray}
For $b_a\tilde{\alpha}_{a}(0)t\gg 1$ we get a fixed point behaviour
in the gauge couplings but not in the gaugino masses:
\begin{eqnarray}
\frac{\tilde{\alpha}_{a}(t)}{\tilde{\alpha}_{c}(t)}
& = & \frac{b_c}{b_a} \nonumber \\
\frac{M_{a}(t)}{M_{c}(t)}
& = & \frac{M_{a}(0)b_c\tilde{\alpha}_c(0)}
{M_{c}(0)b_a\tilde{\alpha}_a(0)}
\end{eqnarray}
Since there is no fixed point to determine the 
gaugino masses we shall introduce the parameters
\begin{eqnarray}
m_{1} & \equiv & 
\frac{M_{1}(0)b_3\tilde{\alpha}_3(0)}
{M_{3}(0)b_1\tilde{\alpha}_1(0)} \nonumber \\
m_{2} & \equiv & \frac{M_{2}(0)b_3\tilde{\alpha}_3(0)}
{M_{3}(0)b_2\tilde{\alpha}_2(0)}
\end{eqnarray}
where $m_1=m_2=1$ corresponds to gaugino unification at the fixed point
(and hence approximate gaugino unification at the intermediate scale.)

Supersymmetry means that
the RGEs for the Yukawa couplings may be factorised into a Yukawa coupling
multiplied by a sum of wavefunction anomalous dimensions for the three legs: 
\begin{eqnarray}
\frac{dY^{h_{ij}}}{dt} &=&Y^{h_{ij}}(N_{Q_{i}}+N_{U_{j}^{c}}+N_{H_{U_{ij}}})
\nonumber \\
\frac{dY^{k_{ij}}}{dt} &=&Y^{k_{ij}}(N_{Q_{i}}+N_{D_{j}^{c}}+N_{H_{D_{ij}}})
\nonumber \\
\frac{dY^{l_{ij}}}{dt} &=&Y^{l_{ij}}(N_{L_{i}}+N_{E_{j}^{c}}+N_{H_{D_{ij}}})
\label{YukRGEs}
\end{eqnarray}
where we have defined $Y^{h_{ij}}\equiv \frac{h_{ij}^{2}}{16\pi ^{2}}$, $%
Y^{k_{ij}}\equiv \frac{k_{ij}^{2}}{16\pi ^{2}}$, $Y^{l_{ij}}\equiv \frac{%
l_{ij}^{2}}{16\pi ^{2}}$. If we assume that the gauge couplings are rapidly
driven to their fixed point ratios then the wavefunction anomalous
dimensions, $N_{i\ }$ may be expressed in terms of the single gauge
coupling $\tilde{\alpha}_{3}$ as: 
\begin{eqnarray}
N_{Q_{i}} &=&(\frac{8}{3}+\frac{3}{2}r_{2}+\frac{1}{30}r_{1})\tilde{\alpha}%
_{3}-\sum_{j=1}^{3}(Y^{h_{ij}}+Y^{k_{ij}})  \nonumber \\
N_{U_{i}^{c}} &=&(\frac{8}{3}+\frac{8}{15}r_{1})\tilde{\alpha}%
_{3}-2\sum_{j=1}^{3}Y^{h_{ji}}  \nonumber \\
N_{D_{i}^{c}} &=&(\frac{8}{3}+\frac{2}{15}r_{1})\tilde{\alpha}%
_{3}-2\sum_{j=1}^{3}Y^{k_{ji}}  \nonumber \\
N_{L_{i}} &=&(\frac{3}{2}r_{2}+\frac{3}{10}r_{1})\tilde{\alpha}%
_{3}-\sum_{j=1}^{3}Y^{l_{ij}}  \nonumber \\
N_{E_{i}^{c}} &=&(\frac{6}{5}r_{1})\tilde{\alpha}_{3}-2%
\sum_{j=1}^{3}Y^{l_{ji}}  \nonumber \\
N_{H_{U_{ij}}} &=&(\frac{3}{2}r_{2}+\frac{3}{10}r_{1})\tilde{\alpha}%
_{3}-3Y^{h_{ij}}  \nonumber \\
N_{H_{D_{ij}}} &=&(\frac{3}{2}r_{2}+\frac{3}{10}r_{1})\tilde{\alpha}%
_{3}-3Y^{k_{ij}}-Y^{l_{ij}}  \label{N}
\end{eqnarray}
The Yukawa RGEs are flavour independent and are driven to the flavour
independent infra-red stable fixed points (IRSFPs) of eq(\ref{YukRGEs}) 
\begin{eqnarray}
{R^{h}}^{*} &=&(\frac{232}{3}+45r_{2}+\frac{232}{15}r_{1}+15b_{3})/219 
\nonumber \\
{R^{k}}^{*} &=&(80+39r_{2}+\frac{21}{15}r_{1}+13b_{3})/219  \nonumber \\
{R^{l}}^{*} &=&(-24+54r_{2}+39r_{1}+18b_{3})/219  \label{FPs}
\end{eqnarray}
where ${R^{h}}^{*}\equiv \frac{{Y^{h}}^{*}}{\tilde{\alpha}_{3}}$, ${R^{k}}%
^{*}\equiv \frac{{Y^{k}}^{*}}{\tilde{\alpha}_{3}}$, ${R^{l}}^{*}\equiv \frac{%
{Y^{l}}^{*}}{\tilde{\alpha}_{3}}$, where ${Y^{h}}^{*}\equiv {Y^{h_{ij}}}^{*}$%
, ${Y^{k}}^{*}\equiv {Y^{k_{ij}}}^{*}$, ${Y^{l}}^{*}\equiv {Y^{l_{ij}}}^{*}$%
, $\forall i,j$. For example for $n=6$ we find $b_{3}=3$, $%
r_{1}=0.238,r_{2}=0.428$, ${R^{h}}^{*}=0.663$, ${R^{k}}^{*}=0.621$, ${R^{l}}%
^{*}=0.285$. (Note the approximate isospin symmetry in the IRSFPs.)

Turning now to the soft SUSY breaking parameters, the trilinear couplings
have RGEs which resemble those of the Yukawa couplings: 
\begin{eqnarray}
\frac{dA_{h_{ij}}}{dt} &=&P_{Q_{i}}+P_{U_{j}^{c}}+P_{H_{U_{ij}}}  \nonumber
\\
\frac{dA_{k_{ij}}}{dt} &=&P_{Q_{i}}+P_{D_{j}^{c}}+P_{H_{D_{ij}}}  \nonumber
\\
\frac{dA_{l_{ij}}}{dt} &=&P_{L_{i}}+P_{E_{j}^{c}}+P_{H_{D_{ij}}}
\label{TrilinearRGEs}
\end{eqnarray}
where 
\begin{eqnarray}
P_{Q_{i}} &=&(\frac{8}{3}+\frac{3}{2}r_{2}m_{2}+\frac{1}{30}r_{1}m_{1})%
\tilde{\alpha}_{3}M_{3}-\sum_{j=1}^{3}(A_{{h_{ij}}}Y^{h_{ij}}+A_{{k_{ij}}%
}Y^{k_{ij}})  \nonumber \\
P_{U_{i}^{c}} &=&(\frac{8}{3}+\frac{8}{15}r_{1}m_{1})\tilde{\alpha}%
_{3}M_{3}-2\sum_{j=1}^{3}A_{h_{ji}}Y^{h_{ji}}  \nonumber \\
P_{D_{i}^{c}} &=&(\frac{8}{3}+\frac{2}{15}r_{1}m_{1})\tilde{\alpha}%
_{3}M_{3}-2\sum_{j=1}^{3}A_{k_{ji}}Y^{k_{ji}}  \nonumber \\
P_{L_{i}} &=&(\frac{3}{2}r_{2}m_{2}+\frac{3}{10}r_{1}m_{1})\tilde{\alpha}%
_{3}M_{3}-\sum_{j=1}^{3}A_{l_{ij}}Y^{l_{ij}}  \nonumber \\
P_{E_{i}^{c}} &=&(\frac{6}{5}r_{1}m_{1})\tilde{\alpha}_{3}M_{3}-2%
\sum_{j=1}^{3}A_{l_{ji}}Y^{l_{ji}}  \nonumber \\
P_{H_{U_{ij}}} &=&(\frac{3}{2}r_{2}m_{2}+\frac{3}{10}r_{1}m_{1})\tilde{\alpha%
}_{3}M_{3}-3A_{h_{ij}}Y^{h_{ij}}  \nonumber \\
P_{H_{D_{ij}}} &=&(\frac{3}{2}r_{2}m_{2}+\frac{3}{10}r_{1}m_{1})\tilde{\alpha%
}_{3}M_{3}-3A_{k_{ij}}Y^{k_{ij}}-A_{l_{ij}}Y^{l_{ij}}  \label{P}
\end{eqnarray}
For illustrative purposes we shall henceforth assume
that $m_{1}=m_{2}=1$ which corresponds to all three gaugino masses
staying equal down to the intermediate scale.  
In this case the IRSFPs for the trilinear couplings
are simply: 
\begin{equation}
A_{h_{ij}}^{*}=A_{k_{ij}}^{*}=A_{l_{ij}}^{*}=M_{3}\ \ (\forall i,j)
\label{TrilinearIRSFPs}
\end{equation}
where $M_{3}=M_{3}(t)$ is the running gaugino mass. This simple result may
be readily understood given the similarity of the trilinear RGEs to the
Yukawa RGEs.

We now turn our attention to the RGEs for the soft scalar 
masses which will be our
principal concern henceforth: 
\begin{eqnarray}
\frac{dm_{Q_{i}}^{2}}{dt} &=&(\frac{16}{3}+3r_{2}m_{2}^{2}+\frac{1}{15}%
r_{1}m_{1}^{2})\tilde{\alpha}_{3}M_{3}^{2}-\sum_{j=1}^{3}(X_{{h_{ij}}%
}Y^{h_{ij}}+X_{{k_{ij}}}Y^{k_{ij}})  \nonumber \\
\frac{dm_{U_{i}^{c}}^{2}}{dt} &=&(\frac{16}{3}+\frac{16}{15}r_{1}m_{1}^{2})%
\tilde{\alpha}_{3}M_{3}^{2}-2\sum_{j=1}^{3}X_{h_{ji}}Y^{h_{ji}}  \nonumber \\
\frac{dm_{D_{i}^{c}}^{2}}{dt} &=&(\frac{16}{3}+\frac{4}{15}r_{1}m_{1}^{2})%
\tilde{\alpha}_{3}M_{3}^{2}-2\sum_{j=1}^{3}X_{k_{ji}}Y^{k_{ji}}  \nonumber \\
\frac{dm_{L_{i}}^{2}}{dt} &=&(3r_{2}m_{2}^{2}+\frac{3}{5}r_{1}m_{1}^{2})%
\tilde{\alpha}_{3}M_{3}^{2}-\sum_{j=1}^{3}X_{l_{ij}}Y^{l_{ij}}  \nonumber \\
\frac{dm_{E_{i}^{c}}^{2}}{dt} &=&(\frac{12}{5}r_{1}m_{1}^{2})\tilde{\alpha}%
_{3}M_{3}^{2}-2\sum_{j=1}^{3}X_{l_{ji}}Y^{l_{ji}}  \nonumber \\
\frac{dm_{H_{U_{ij}}}^{2}}{dt} &=&(3r_{2}m_{2}^{2}+\frac{3}{5}r_{1}m_{1}^{2})%
\tilde{\alpha}_{3}M_{3}^{2}-3X_{h_{ij}}Y^{h_{ij}}  \nonumber \\
\frac{dm_{H_{D_{ij}}}^{2}}{dt} &=&(3r_{2}m_{2}^{2}+\frac{3}{5}r_{1}m_{1}^{2})%
\tilde{\alpha}_{3}M_{3}^{2}-3X_{k_{ij}}Y^{k_{ij}}-X_{l_{ij}}Y^{l_{ij}}
\label{mRGEs}
\end{eqnarray}
where 
\begin{eqnarray}
X_{{h_{ij}}}
&=&m_{Q_{i}}^{2}+m_{U_{j}^{c}}^{2}+m_{H_{U_{ij}}}^{2}+A_{h_{ij}}^{2} 
\nonumber \\
X_{{k_{ij}}}
&=&m_{Q_{i}}^{2}+m_{D_{j}^{c}}^{2}+m_{H_{D_{ij}}}^{2}+A_{k_{ij}}^{2} 
\nonumber \\
X_{{l_{ij}}}
&=&m_{L_{i}}^{2}+m_{E_{j}^{c}}^{2}+m_{H_{D_{ij}}}^{2}+A_{l_{ij}}^{2}
\label{X}
\end{eqnarray}

It is straightforward to see that Eq\ref{mRGEs} have IRSFPs given by 
\begin{equation}
X_{h_{ij}}^{*}=X_{k_{ij}}^{*}=X_{l_{ij}}^{*}=2M_{3}^{2}\ \ (\forall i,j)
\label{XIRSFPs}
\end{equation}
At first sight a result like that in Eq.\ref{XIRSFPs} seems very exciting
since it would appear to provide a prediction for the soft SUSY\ breaking
masses. However it also involves the unknown (and unmeasurable since the
additional Higgs fields are very heavy) soft mass parameters of all the 18
Higgs doublets.

To determine the implications for the individual masses we will solve the
RGEs for the soft scalar masses in the simplified case where $m_{1}=m_{2}=1$,
inserting the IRSFPs for the gauge couplings in Eq.\ref{r}, and the IRSFPs
for the Yukawa couplings and trilinear couplings in Eqs.\ref{FPs}, \ref
{TrilinearIRSFPs}. With these assumptions the RGEs for the soft masses
become: 
\begin{eqnarray}
\frac{dm_{Q_{i}}^{2}}{dt} &=&(\frac{16}{3}+3r_{2}+\frac{1}{15}r_{1})\tilde{%
\alpha}_{3}M_{3}^{2}-{Y^{h}}%
^{*}(3m_{Q_{i}}^{2}+m_{U_{T}^{c}}^{2}+m_{H_{U_{R_{i}}}}^{2}+3M_{3}^{2}) 
\nonumber  \label{mIRSFPRGEs} \\
&-&{Y^{k}}%
^{*}(3m_{Q_{i}}^{2}+m_{D_{T}^{c}}^{2}+m_{H_{D_{R_{i}}}}^{2}+3M_{3}^{2}) 
\nonumber \\
\frac{dm_{U_{i}^{c}}^{2}}{dt} &=&(\frac{16}{3}+\frac{16}{15}r_{1})\tilde{%
\alpha}_{3}M_{3}^{2}-2{Y^{h}}%
^{*}(m_{Q_{T}}^{2}+3m_{U_{i}^{c}}^{2}+m_{H_{U_{C_{i}}}}^{2}+3M_{3}^{2}) 
\nonumber \\
\frac{dm_{D_{i}^{c}}^{2}}{dt} &=&(\frac{16}{3}+\frac{4}{15}r_{1})\tilde{%
\alpha}_{3}M_{3}^{2}-2{Y^{k}}%
^{*}(m_{Q_{T}}^{2}+3m_{D_{i}^{c}}^{2}+m_{H_{D_{C_{i}}}}^{2}+3M_{3}^{2}) 
\nonumber \\
\frac{dm_{L_{i}}^{2}}{dt} &=&(3r_{2}+\frac{3}{5}r_{1})\tilde{\alpha}%
_{3}M_{3}^{2}-{Y^{l}}%
^{*}(3m_{L_{i}}^{2}+m_{E_{T}^{c}}^{2}+m_{H_{D_{R_{i}}}}^{2}+3M_{3}^{2}) 
\nonumber \\
\frac{dm_{E_{i}^{c}}^{2}}{dt} &=&(\frac{12}{5}r_{1})\tilde{\alpha}%
_{3}M_{3}^{2}-2{Y^{l}}%
^{*}(m_{L_{T}}^{2}+3m_{E_{i}^{c}}^{2}+m_{H_{D_{C_{i}}}}^{2}+3M_{3}^{2}) 
\nonumber \\
\frac{dm_{H_{U_{R_{i}}}}^{2}}{dt} &=&3(3r_{2}+\frac{3}{5}r_{1})\tilde{\alpha}%
_{3}M_{3}^{2}-3{Y^{h}}%
^{*}(3m_{Q_{i}}^{2}+m_{U_{T}^{c}}^{2}+m_{H_{U_{R_{i}}}}^{2}+3M_{3}^{2}) 
\nonumber \\
\frac{dm_{H_{U_{C_{i}}}}^{2}}{dt} &=&3(3r_{2}+\frac{3}{5}r_{1})\tilde{\alpha}%
_{3}M_{3}^{2}-3{Y^{h}}%
^{*}(m_{Q_{T}}^{2}+3m_{U_{i}^{c}}^{2}+m_{H_{U_{C_{i}}}}^{2}+3M_{3}^{2}) 
\nonumber \\
\frac{dm_{H_{D_{R_{i}}}}^{2}}{dt} &=&3(3r_{2}+\frac{3}{5}r_{1})\tilde{\alpha}%
_{3}M_{3}^{2}-3{Y^{k}}%
^{*}(3m_{Q_{i}}^{2}+m_{D_{T}^{c}}^{2}+m_{H_{D_{R_{i}}}}^{2}+3M_{3}^{2}) 
\nonumber \\
&-&{Y^{l}}%
^{*}(3m_{L_{i}}^{2}+m_{E_{T}^{c}}^{2}+m_{H_{D_{R_{i}}}}^{2}+3M_{3}^{2}) 
\nonumber \\
\frac{dm_{H_{D_{C_{i}}}}^{2}}{dt} &=&3(3r_{2}+\frac{3}{5}r_{1})\tilde{\alpha}%
_{3}M_{3}^{2}-3{Y^{k}}%
^{*}(m_{Q_{T}}^{2}+3m_{D_{i}^{c}}^{2}+m_{H_{D_{C_{i}}}}^{2}+3M_{3}^{2}) 
\nonumber \\
&-&{Y^{l}}%
^{*}(m_{L_{T}}^{2}+3m_{E_{i}^{c}}^{2}+m_{H_{D_{C_{i}}}}^{2}+3M_{3}^{2}) 
\nonumber \\
&&
\label{rge1}
\end{eqnarray}
where subscript $T$ indicates the sum over families of the squark and
slepton masses, 
\begin{eqnarray}
m_{Q_{T}}^{2} &=&m_{Q_{1}}^{2}+m_{Q_{2}}^{2}+m_{Q_{3}}^{2}  \nonumber \\
m_{U_{T}^{c}}^{2} &=&m_{U_{1}^{c}}^{2}+m_{U_{2}^{c}}^{2}+m_{U_{3}^{c}}^{2} 
\nonumber \\
m_{D_{T}^{c}}^{2} &=&m_{D_{1}^{c}}^{2}+m_{D_{2}^{c}}^{2}+m_{D_{3}^{c}}^{2} 
\nonumber \\
m_{L_{T}}^{2} &=&m_{L_{1}}^{2}+m_{L_{2}}^{2}+m_{L_{3}}^{2}  \nonumber \\
m_{E_{T}^{c}}^{2} &=&m_{E_{1}^{c}}^{2}+m_{E_{2}^{c}}^{2}+m_{E_{3}^{c}}^{2}
\label{T}
\end{eqnarray}
and subscript $R_{i}$ and $C_{i}$ indicates the sum of the ith row and
column of the matrix of Higgs masses, for both $H_{U}$ and $H_{D}$, 
\begin{eqnarray}
m_{H_{U_{R_{i}}}}^{2}
&=&m_{H_{U_{i1}}}^{2}+m_{H_{U_{i2}}}^{2}+m_{H_{U_{i3}}}^{2}  \nonumber \\
m_{H_{U_{C_{i}}}}^{2}
&=&m_{H_{U_{1i}}}^{2}+m_{H_{U_{2i}}}^{2}+m_{H_{U_{3i}}}^{2}  \nonumber \\
m_{H_{D_{R_{i}}}}^{2}
&=&m_{H_{D_{i1}}}^{2}+m_{H_{D_{i2}}}^{2}+m_{H_{D_{i3}}}^{2}  \nonumber \\
m_{H_{D_{C_{i}}}}^{2}
&=&m_{H_{D_{1i}}}^{2}+m_{H_{D_{2i}}}^{2}+m_{H_{D_{3i}}}^{2}  \label{RC}
\end{eqnarray}
It is also convenient to define the total sum of Higgs masses, 
\begin{eqnarray}
m_{H_{U_{T}}}^2 &=&\sum_{i,j=1}^{3}m_{H_{U_{ij}}}^{2}  \nonumber \\
m_{H_{D_{T}}}^2 &=&\sum_{i,j=1}^{3}m_{H_{D_{ij}}}^{2}
\end{eqnarray}

\section{Fixed point structure for family sums of soft scalar masses}

In this section we shall consider the fixed points obtained
by summing Eq.\ref{mIRSFPRGEs} over flavours. In this way one obtains
the coupled set of equations for the $T$ mass variables:
\begin{eqnarray}
\frac{dm_{Q_T}^2}{dt} &=&3(\frac{16}{3} + 3r_2+ \frac{1}{15} r_1)
\tilde{\alpha}_3 M_3^2
-  {Y^h}^*(\bar{m}_U^2 +9M_3^2)  - {Y^k}^* (\bar{m}_D^2+9M_3^2) 
\nonumber \\
\frac{dm_{U_T^c}^2}{dt} &=&3(\frac{16}{3}+\frac{16}{15}r_{1})
\tilde{\alpha}_{3}M_{3}^{2}-2{Y^{h}}^{*}(\bar{m}_U^2 +9M_{3}^{2}) 
\nonumber \\
\frac{dm_{D_T^c}^2 }{dt} &=&3(\frac{16}{3}+\frac{4}{15}r_{1})
\tilde{\alpha}_{3}M_{3}^{2}-2{Y^{k}}^{*}(\bar{m}_D^2 +9M_{3}^{2}) 
\nonumber \\
\frac{dm_{L_T}^2 }{dt} &=&3(3r_{2}+\frac{3}{5}r_{1})
\tilde{\alpha}_{3}M_{3}^{2}-{Y^{l}}^{*}(\bar{m}_E^2+9M_{3}^{2}) 
\nonumber \\
\frac{dm_{E_T^c}^2 }{dt} &=&3(\frac{12}{5}r_{1})
\tilde{\alpha}_{3}M_{3}^{2}-2{Y^{l}}^{*}(\bar{m}_E^2+9M_{3}^{2}) 
\nonumber \\
\frac{dm_{H_{U_T}}^2 }{dt} &=&9(3r_{2}+\frac{3}{5}r_{1})
\tilde{\alpha}_{3}M_{3}^{2}-3{Y^{h}}^{*}(\bar{m}_U^2+9M_{3}^{2}) 
\nonumber \\
\frac{dm_{H_{D_T}}^2 }{dt} &=&9(3r_{2}+\frac{3}{5}r_{1})
\tilde{\alpha}_{3}M_{3}^{2}
- 3{Y^{k}}^{*}(\bar{m}_D^2+9M_{3}^{2})  
-{Y^{l}}^{*}(\bar{m}_E^2 +9M_{3}^{2}) 
\label{mTeqs}
\end{eqnarray}
where we have defined the combinations
\begin{eqnarray}
\bar{m}_U^2 & = & 3m_{Q_T}^2+3m_{U_T^c}^2+m_{H_{U_T}}^2 \nonumber \\
\bar{m}_D^2 & = & 3m_{Q_T}^2+3m_{D_T^c}^2+m_{H_{D_T}}^2 \nonumber \\
\bar{m}_E^2 & = & 3m_{L_T}^2+3m_{E_T^c}^2+m_{H_{D_T}}^2
\label{mbar}
\end{eqnarray}

Note that the right hand sides of the RGEs in Eq.\ref{mTeqs}
only involve the 3 independent
combinations of soft scalar masses in Eq.\ref{mbar}, 
which satisfy the following coupled equations:
\begin{eqnarray}
\frac{d\bar{m}_U^2  }{dt} &=&9(\frac{32}{3} + 6r_2+ \frac{26}{15} r_1)
\tilde{\alpha}_3 M_3^2
- 12 {Y^h}^*(\bar{m}_U^2 +9M_3^2)  - 3{Y^k}^* (\bar{m}_D^2+9M_3^2) 
\nonumber \\
\frac{d\bar{m}_D^2 }{dt} &=&9(\frac{32}{3}+ 6r_2 + \frac{14}{15}r_{1})
\tilde{\alpha}_{3}M_{3}^{2}-3{Y^{h}}^{*}(\bar{m}_U^2 +9M_{3}^{2}) 
-12{Y^{k}}^{*}(\bar{m}_D^2 +9M_{3}^{2})
\nonumber \\
&-&{Y^{l}}^{*}(\bar{m}_E^2 +9M_{3}^{2})  
\nonumber \\
\frac{d\bar{m}_E^2 }{dt} &=&9(6r_2 + \frac{18}{5}r_{1})
\tilde{\alpha}_{3}M_{3}^{2}
-3{Y^{k}}^{*}(\bar{m}_D^2 +9M_{3}^{2})-10{Y^{l}}^{*}(\bar{m}_E^2 +9M_{3}^{2})  
\label{nonzeromodes}
\end{eqnarray}
The coupled RGEs in Eq.\ref{nonzeromodes} may be expressed in terms of a
symmetric matrix (after appropriate normalisation factors are included)
which may then be diagonalised.
In the diagonal basis these 3 RGEs then have the form:
\begin{equation}
\frac{d\bar{m}_{T_{i}}^{2}}{dt}
=-A_{i}\tilde{\alpha}_{3}\bar{m}_{T_{i}}^{2}+B_{i}
\tilde{\alpha}_{3}M_{3}^{2}  \label{mTRGE}
\end{equation}
with the $A_{i},B_{i}$ coefficients given in Table 1. 
The equations may then be
solved to yield the low energy soft total masses: 
\begin{equation}
\bar{m}_{T_{i}}^{2}(t)=\bar{m}_{T_{i}}^{2}(0)
\left( \frac{\tilde{\alpha}_{3}(t)}{\tilde{%
\alpha}_{3}(0)}\right) ^{\frac{A_{i}}{b_{3}}}+\frac{B_{i}}{b_{3}(2-\frac{%
A_{i}}{b_{3}})}M_{3}^{2}(0)\left[ \left( \frac{\tilde{\alpha}_{3}(t)}{\tilde{%
\alpha}_{3}(0)}\right) ^{\frac{A_{i}}{b_{3}}}-\left( \frac{\tilde{\alpha}%
_{3}(t)}{\tilde{\alpha}_{3}(0)}\right) ^{2}\right]  \label{mTsoln}
\end{equation}

The orthogonal combinations of soft $T$ masses satisfy RGEs which do not
depend on the soft masses themselves. The combinations are 
not unique due to the degeneracy, but a simple choice is:
\begin{eqnarray}
\frac{d(2m_{Q_T}^2-m_{U_T^c}^2 -m_{D_T^c}^2 )}{dt} 
&=&3(6r_2 - \frac{19}{15}r_{1})\tilde{\alpha}_{3}M_{3}^{2}
\nonumber \\
\frac{d(2m_{L_T}^2-m_{E_T^c}^2) }{dt} 
&=&3(6r_{2}-\frac{6}{5}r_{1})\tilde{\alpha}_{3}M_{3}^{2}
\nonumber \\
\frac{d(3m_{U_T^c}^2-2m_{H_{U_T}}^2  ) }{dt} 
&=&9(\frac{16}{3} - 6r_{2}-\frac{2}{15}r_{1})\tilde{\alpha}_{3}M_{3}^{2}
\nonumber \\
\frac{d(3m_{D_T^c}^2+m_{E_T^c}^2-2m_{H_{D_T}}^2  ) }{dt} 
&=&9(\frac{16}{3} - 6r_{2}-\frac{2}{15}r_{1})\tilde{\alpha}_{3}M_{3}^{2}
\label{zeromodes}
\end{eqnarray}
One may see that eqs.\ref{zeromodes} have a simpler form with $A_i=0$,
\begin{equation}
\frac{dm_{T_{i}}^{2}}{dt}=B_{i}\tilde{\alpha}_{3}M_{3}^{2}  \label{mTRGE0}
\end{equation}
where the $m_{T_{i}}$ and $B_i$ for $i=4\cdots 7$ may be read from
Eq. \ref{zeromodes}.
The low energy solutions in this case are
simply obtained by setting $A_{i}=0$ in Eq.\ref{mTsoln}:
\begin{equation}
m_{T_{i}}^{2}(t)=
m_{T_{i}}^{2}(0)+\frac{B_{i}}{2b_{3}}[M_{3}^{2}(0)-M_{3}^{2}(t)]  
\approx m_{T_{i}}^{2}(0)+\frac{B_{i}}{2b_{3}}M_{3}^{2}(0)
\label{zero}
\end{equation}
where the last approximation, $M_{3}^{2}(0)\gg M_{3}^{2}(t)$, follows 
since  in strong unification theories
the gaugino masses have
radiative corrections which cause them to run in proportion to the gauge
couplings. 

\begin{table}[tbp]
\hfil
\begin{tabular}{ccccccc}
\hline
$n$ & $A_1$ & $B_1$ & $A_2$ & $B_2$ & $A_3$ & $B_3$ \\ \hline
 6  & 9.7   & 7.1   & 5.9   & 0.024 & 2.7   & 2.2   \\ \hline
 8  & 12.1  & 4.0   & 7.4   & 0.72  & 4.7   & 3.2    \\ \hline
10  & 14.3  & 0.65  & 9.0   & 1.8   & 6.3   & 3.9    \\ \hline
20  & 24.9  & 18.2  & 17.3  & -7.3  & 12.9  & 4.5     \\ \hline
40  & 45.1  & 58.3  & 34.1  & -12.9 & 24.0  & 4.2     \\ \hline
\end{tabular}
\hfil
\caption{$A_{i}$ and $B_{i}$ coefficients as a function of $n$.}
\end{table}

It is clear that the linear combinations of soft scalar masses in 
Eq.\ref{zeromodes} will have low energy values in Eq.\ref{zero}
of order the high energy boundary
values of soft scalar masses and gaugino masses at $t=0$, 
while the combinations of soft scalar masses in Eq.\ref{mbar}
will have values at low energy in Eq.\ref{mTsoln}
suppressed by 
$\left( \frac{\tilde{\alpha}_{3}(t)}{\tilde{\alpha}_{3}(0)}\right) 
^{\frac{A_{i}}{b_{3}}}$ 
where the anomalous dimension $\frac{A_{i}}{b_{3}}
=A_{i}/(n-3)\stackrel{>}{\sim } 1$ 
according to Table 1\footnote{
Note that our model of masses needs at least 16 additional Higgs
doublets plus their vector partners. If these all orginate from 
$5+ \bar{5}$ then we need $n_5 \geq 16$ and hence we need
$n\geq 16$.}.
Thus in strong unification where the ratio of gauge couplings is large
it is reasonable to neglect the 
linear combinations of soft scalar masses which are suppressed by anomalous
dimensions compared to the linear combinations of soft scalar masses 
which are unsuppressed by anomalous dimensions, and so at low energy we have
\begin{eqnarray}
3m_{Q_T}^2(t)+3m_{U_T^c}^2(t)+m_{H_{U_T}}^2(t) &\approx 0& \nonumber \\
3m_{Q_T}^2(t)+3m_{D_T^c}^2(t)+m_{H_{D_T}}^2(t) &\approx 0& \nonumber \\
3m_{L_T}^2(t)+3m_{E_T^c}^2(t)+m_{H_{D_T}}^2(t) &\approx 0&
\label{mapprox}
\end{eqnarray}
The fact that these combinations have been shown to be suppressed by
anomalous dimensions is more or less equivalent to the statement
that the combinations in Eq.\ref{X} are fixed points. The more or less
part is due to the fact that one must sum Eq.\ref{X} over flavours,
insert the trilinear fixed point, and then observe that the 
anomalous suppression of these mass combinations is competitive
to that of the gaugino mass.

Turning now to the orthogonal combinations of soft scalar masses,
the appearance of combinations of soft masses whose RGEs do 
not depend on the soft masses themselves
means that the flow to IRSFPs does not lead to suppression
by anomalous dimensions, and so
the low-energy values depend sensitively on the initial values
as seen in Eq.\ref{zero}.
This is in contrast to the fixed point
structure of the Yukawa couplings which do drive the Yukawa couplings to
family independent fixed points. 
Combining Eq.\ref{zero} with Eq.\ref{mapprox} we may solve for the
individual soft scalar mass sums and find 
\begin{eqnarray} 
m_{Q_{T}}^{2}(t) &=&c_{Q}M_{3}^{2}(0)  
+ {\displaystyle \frac {87}{146}} \,{\it m_{Q_{T}}^{2}(0) } 
- {\displaystyle \frac {29}{146}} \,{\it m_{U_{T}^{c}}^{2}(0) }
 - {\displaystyle \frac {15}{73}} \,{\it m_{D_{T}^{c}}^{2}(0) } 
\nonumber \\ & & \mbox{} 
+ {\displaystyle \frac {3}{146}} \,{\it m_{L_{T}}^{2}(0)} 
+ {\displaystyle \frac {3}{146}} \,{\it m_{E_{T}^{c}}^{2}(0) } 
- {\displaystyle \frac {29}{438}} \,{\it m_{H_{U_T}}^2(0) } 
- {\displaystyle \frac {9}{146}} \,{\it m_{H_{D_T}}^2(0) }
\nonumber \\
m_{U_{T}^{c}}^{2}(t) &=&c_{U}M_{3}^{2}(0)  
- {\displaystyle \frac {29}{73}} \,{\it m_{Q_{T}}^{2}(0) } 
+ {\displaystyle \frac {34}{73}} \,{\it m_{U_{T}^{c}}^{2}(0) } 
+ {\displaystyle \frac {10}{73}} \,{\it m_{D_{T}^{c}}^{2}(0) } 
\nonumber \\ & & \mbox{} 
- {\displaystyle \frac {1}{73}} \,{\it m_{L_{T}}^{2}(0)} 
- {\displaystyle \frac {1}{73}} \,{\it m_{E_{T}^{c}}^{2}(0) } 
- {\displaystyle \frac {13}{73}} \,{\it m_{H_{U_T}}^2(0) }
+ {\displaystyle \frac {3}{73}} \,{\it m_{H_{D_T}}^2(0) }
\nonumber \\ 
m_{D_{T}^{c}}^{2}(t) &=&c_{D}M_{3}^{2}(0)  
- {\displaystyle \frac {30}{73}} \,{\it m_{Q_{T}}^{2}(0) } 
+ {\displaystyle \frac {10}{73}} \,{\it m_{U_{T}^{c}}^{2}(0) } 
+ {\displaystyle \frac {33}{73}} \,{\it m_{D_{T}^{c}}^{2}(0) } 
\nonumber \\ & & \mbox{} 
+ {\displaystyle \frac {4}{73}} \,{\it m_{L_{T}}^{2}(0)} 
+ {\displaystyle \frac {4}{73}} \,{\it m_{E_{T}^{c}}^{2}(0) } 
+ {\displaystyle \frac {10}{219}} \,{\it m_{H_{U_T}}^2(0) }
- {\displaystyle \frac {12}{73}} \,{\it m_{H_{D_T}}^2(0) }
\nonumber \\ 
m_{L_{T}}^{2}(t) &=&c_{L}M_{3}^{2}(0)  
+ {\displaystyle \frac {9}{146}} \,{\it m_{Q_{T}}^{2}(0) } 
- {\displaystyle \frac {3}{146}} \,{\it m_{U_{T}^{c}}^{2}(0) } 
+ {\displaystyle \frac {6}{73}} \,{\it m_{D_{T}^{c}}^{2}(0) } 
\nonumber \\ & & \mbox{} 
+ {\displaystyle \frac {101}{146}} \,{\it m_{L_{T}}^{2}(0)} 
- {\displaystyle \frac {45}{146}} \,{\it m_{E_{T}^{c}}^{2}(0) } 
- {\displaystyle \frac {1}{146}} \,{\it m_{H_{U_T}}^2(0) } 
- {\displaystyle \frac {11}{146}} \,{\it m_{H_{D_T}}^2(0) }
\nonumber \\
m_{E_{T}^{c}}^{2}(t) &=&c_{E}M_{3}^{2}(0)  
+ {\displaystyle \frac {9}{73}} \,{\it m_{Q_{T}}^{2}(0) } 
- {\displaystyle \frac {3}{73}} \,{\it m_{U_{T}^{c}}^{2}(0) } 
+ {\displaystyle \frac {12}{73}} \,{\it m_{D_{T}^{c}}^{2}(0) } 
\nonumber \\ & & \mbox{} 
- {\displaystyle \frac {45}{73}} \,{\it m_{L_{T}}^{2}(0)} 
+ {\displaystyle \frac {28}{73}} \,{\it m_{E_{T}^{c}}^{2}(0) } 
- {\displaystyle \frac {1}{73}} \,{\it m_{H_{U_T}}^2(0) } 
- {\displaystyle \frac {11}{73}} \,{\it m_{H_{D_T}}^2(0) }
\nonumber \\
m_{H_{U_T}}^2(t) &=&c_{H_U}M_{3}^{2}(0)  
- {\displaystyle \frac {87}{146}} \,{\it m_{Q_{T}}^{2}(0) } 
- {\displaystyle \frac {117}{146}} \,{\it m_{U_{T}^{c}}^{2}(0) } 
+ {\displaystyle \frac {15}{73}} \,{\it m_{D_{T}^{c}}^{2}(0) } 
\nonumber \\ & & \mbox{} 
- {\displaystyle \frac {3}{146}} \,{\it m_{L_{T}}^{2}(0)} 
- {\displaystyle \frac {3}{146}} \,{\it m_{E_{T}^{c}}^{2}(0) } 
+ {\displaystyle \frac {107}{146}} \,{\it m_{H_{U_T}}^2(0) } 
+ {\displaystyle \frac {9}{146}} \,{\it m_{H_{D_T}}^2(0) }
\nonumber \\
m_{H_{D_T}}^2(t) &=&c_{H_D}M_{3}^{2}(0)
- {\displaystyle \frac {81}{146}} \,{\it m_{Q_{T}}^{2}(0) } 
+ {\displaystyle \frac {27}{146}} \,{\it m_{U_{T}^{c}}^{2}(0) }
 - {\displaystyle \frac {54}{73}} \,{\it m_{D_{T}^{c}}^{2}(0) } 
\nonumber \\ & & \mbox{} 
- {\displaystyle \frac {33}{146}} \,{\it m_{L_{T}}^{2}(0)} 
- {\displaystyle \frac {33}{146}} \,{\it m_{E_{T}^{c}}^{2}(0) } 
+ {\displaystyle \frac {9}{146}} \,{\it m_{H_{U_T}}^2(0) } 
+ {\displaystyle \frac {99}{146}} \,{\it m_{H_{D_T}}^2(0) }
\label{fullmTpredictions}
\end{eqnarray}
where the predictions are valid at the scale $M_{I}$ and the coefficients 
$c_{Q},c_{U},c_{D},c_{L},c_{E},c_{H_U},c_{H_D}$  are given in Table 2. 
The coefficients, $c_i$, satisfy the following sum rules
which follow from Eq.\ref{mapprox}:
\begin{eqnarray}
3c_Q+3c_U+c_{H_U} & = & 0, \nonumber \\
3c_Q+3c_D+c_{H_D} & = & 0, \nonumber \\
3c_L+3c_E+c_{H_D}& = & 0. \nonumber \\
\label{sumrules}
\end{eqnarray}

An attractive possibility is that the soft scalar masses at high
energy may be neglected compared to the gaugino masses at high energy and
thus the dominant contribution to the low-energy masses is from the family
independent couplings to the gaugino sector (a supergravity version of gauge
mediation). In this case we may neglect $m_{T_{i}}^{2}(0)$
compared to $M_{3}^{2}(0)$, leading to the simple low energy predictions
\begin{eqnarray}
m_{Q_{T}}^{2}(t) &\approx &c_{Q}M_{3}^{2}(0)  \nonumber \\
m_{U_{T}^{c}}^{2}(t) &\approx &c_{U}M_{3}^{2}(0)  \nonumber \\
m_{D_{T}^{c}}^{2}(t) &\approx &c_{D}M_{3}^{2}(0)  \nonumber \\
m_{L_{T}}^{2}(t) &\approx &c_{L}M_{3}^{2}(0)  \nonumber \\
m_{E_{T}^{c}}^{2}(t) &\approx &c_{E}M_{3}^{2}(0)  \nonumber \\
m_{H_{U_T}}^2(t) &\approx &c_{H_U}M_{3}^{2}(0)  \nonumber \\
m_{H_{D_T}}^2(t) &\approx &c_{H_D}M_{3}^{2}(0)
\label{mTpredictions}
\end{eqnarray}

Notice that the
coefficients $c_{U},c_{D},c_{E},c_{H_U},c_{H_D}$ 
are driven negative. The reason is simply that the Yukawa couplings 
which control the negative contributions to the RGEs in Eq.\ref{mTeqs}
are of the same order as the gauge couplings (related by the fixed point),
so soft scalar 
masses with small gauge Casimirs and/or large Yukawa multiplicities
are prone to be driven negative. Although we have unearthed this problem in the
gauge dominated approximation, the result follows from the structure of 
the RGEs and applies to more general initial conditions.

Clearly it is unacceptable to have squark or slepton masses negative. However 
there are two possible reasons why this result 
need not lead us to abandon the underlying model. 
The first possibility is that the theory does 
not have time to settle in to its fixed point. 
The evolution of the couplings is cut-off at the scale $M_I$, the 
superpotential mass term associated with the heavy states. This term is 
generated through couplings of the vectorlike fields $X_I$, $\bar{X}_I$ to a 
Standard Model singlet field, $\Phi$, 
via a coupling in the superpotential of the form  $\lambda 
\bar{X}_I X_I \Phi$ where, for simplicity,  we have taken the couplings to 
different components to be equal at the unification scale; this could arise 
because of an enhanced symmetry 
at this scale. The soft mass squared of the scalar component of the gauge 
singlet field $\Phi$ will be driven negative through the terms in the 
renormalisation group equations involving the Yukawa coupling 
$\lambda$ - there 
are no stabilising terms proportional to the 
gaugino mass squared because $\Phi$ is a gauge singlet field.  As 
a result a singlet vev will be generated. 
Its magnitude depends on how flat the 
$\Phi$ potential is. In the extreme, when there are no $\Phi$ F-terms, the 
$\Phi$ vev will be close to the scale at which the 
$\Phi$ mass squared becomes negative.  The 
details of this depend on the initial value of the $\Phi$ soft mass but, given 
that $\Phi$ has no stabilising D-terms, it is possible that this will occur 
before the squark mass
squared becomes negative. 
Since the $\Phi$ vev generates the mass for the heavy 
states, $M_I=\lambda <\Phi >$, the contribution of the heavy states to the 
renormalisation group evolution drops out at this scale. 
Thus the terms driving 
the squark mass squared negative cease to be 
effective before the mass squared actually becomes negative, 
i.e. the fixed point discussed above is not reached. In this case the 
phenomenological implications for the soft 
masses are not correctly captured by 
Eqs.\ref{mTpredictions}. 
Instead, to a first approximation, the masses will simply be proportional 
to the coefficient of the gaugino mass squared term in Eqs.\ref{rge1} coming 
from the one loop term coupling the squarks to the gauginos. 
Note that in strong unification this
gives a characteristically different pattern of 
masses to that found in the MSSM under the 
same assumption of gaugino mass domination. 
The reason is because the three gauge couplings 
are nearly the same in strong unification close 
to the unification scale and hence the differences 
between the contribution to the soft masses of the 
various gauge interactions is reduced. 

The case we have just discussed loses many of the attractive features we have 
been seeking to explore because there is not enough time to reach the fixed 
points. In this case the predictive power of the fixed point structure is 
largely lost, the low-energy values being 
sensitive to the initial conditions. 
Thus we turn to a consideration of the 
second possible way to avoid the appearance 
of negative mass squared terms for the squark and slepton masses. 
This follows because their appearance is a model dependent 
result dependent on the structure assumed for the Yukawa 
couplings in the heavy sector. As we shall discuss 
inclusion of additional couplings may leave all 
squark and slepton masses positive. In this case the fixed point structure is 
phenomenologically viable.
Provided the value of $M_I$ is low enough for the fixed points to be reached 
the dependence on the precise value of $M_I$ and on other initial conditions 
will be small and one may realise the predictive power of the infra-red 
structure. 
In the next section we discuss why it is 
possible that the negative eigenvalues we found for the squark and slepton 
masses squared arose due to an 
oversimplification of the model. 
\begin{table}[tbp]
\hfil
\begin{tabular}{cccccccc}
\hline
$n$ & $c_Q$ & $c_U$ & $c_D$ & $c_L$ & $c_E$ & $c_{H_U}$ & $c_{H_D}$\\ \hline
6 & 0.61 & 0.047 & 0.041 & 0.60 & 0.054 & -1.98 & -1.96\\ \hline
8 & 0.38 & -0.058 & -0.050 & 0.40 & -0.07 & -0.97 & -0.99 \\ \hline
10 & 0.28 & -0.08 & -0.069 & 0.31 & -0.10 & -0.59 & -0.62 \\ \hline
20 & 0.12 & -0.066 & -0.057 & 0.14 & -0.081 & -0.15 &  -0.18\\ \hline
40 & 0.054 & -0.038 & -0.033 & 0.067 & -0.047 & -0.046 & -0.061\\ \hline
\end{tabular}
\hfil
\caption{$c_Q,c_U,c_D,c_L,c_E,c_{H_U},c_{H_D}$ 
coefficients as a function of $n$.}
\end{table}

\section{Extra Higgs Couplings}

The results in the previous section highlighted a problem
with the fixed point structure of soft masses, namely that the squark and 
slepton scalars
tend to have their mass squared driven negative by the large Yukawa
couplings.
The same effect also applies to the Higgs masses, although in this
case this does not necessarily imply that vevs for the Higgs will result.
The point is that the Higgs masses which enter the effective potential
will receive an additional contribution from their supersymmetric 
mass terms,  the analogue of the
$\mu$ parameter of the MSSM, and it is the combination of the form
$\mu^2 + m^2$ which is relevant for symmetry breaking, although only
the Higgs soft mass squared $m^2$ enters the RGEs for the squark and slepton
soft masses \footnote{ Of course there are also supersymmetric 
mass terms for the squarks and sleptons 
which need to be included, but for the light generations 
these will be too small to counteract the effect of a 
negative soft scalar mass squared.}. This might seem rather 
irrelevant since in any case the
squark and slepton mass squared get driven negative, but as we shall
see it is rather easy to modify the model so that only the Higgs mass
squareds get driven negative, and our point is then that the modified model
is viable. 

The Higgs sector of the model in ref.\cite{GG}
is more complicated than that so far considered and, 
as we have discussed, includes
two classes of Standard Model gauge group singlets 
which couple to the Higgs  : the $\theta$, $\bar{\theta}$
singlets which couple a Higgs of a 
given $X$ charge to another Higgs with whose $X$ charge differs by one unit;
and $\Phi$ singlets which carry no $X$ charge and couple pairs of Higgs
fields with equal and opposite $X$ charge, so that the VEV of $\Phi$
essentially generates the intermediate scale $M_I$.
So far we have not considered the effect of either of these two types
of additional coupling in the Higgs sector, although their effect on the
Yukawa sector has already been considered \cite{GG}, \cite{AK}.
We have been implicitly assuming that the extra singlet couplings
do not significantly disrupt the flavour symmetry of the Yukawa couplings,
which has some support from the analysis in \cite{GG}.
However the presence of these couplings will generate additional Higgs
masses in the superpotential, analagous to the $\mu$ parameter of the MSSM.
For all the Higgs with non-zero $X$ charge
these generalised $\mu$ parameters will all be of order the intermediate
scale $M_I$, but for the Higgs $H_{U_{33}}$, $H_{D_{33}}$ which do
not have vector partners, and hence do not couple to the $\Phi$ singlets,
there is no such intermediate scale mass contribution.
However, as already mentioned, these Higgs
give the dominant component of the two Higgs doublets of the MSSM,
and we know that for a viable phenomenology 
there must be a mechanism which couples them 
together to form a conventional $\mu$ term. 
In the case under consideration the $\mu$ term 
that is required must be sufficient
to overcome the negative mass squared generated at the intermediate
scale by radiative corrections and so must be at least of order $M_3(0)$
which implies a rather large value for the effective $\mu$ parameter at the
intermediate scale. 

Another effect of the extra Higgs Yukawa couplings is to drive the
Higgs masses even more negative at the intermediate scale than our
previous estimate suggests. This is a generic expectation based on the
fact that extra large Yukawa couplings will always tend to drive a 
soft scalar mass squared negative more quickly. As far as the Higgs masses are
concerned this is not a problem for the reasons outlined above.
On the contrary this effect is actually welcome since the more quickly
the Higgs mass squareds are driven negative, the less quickly will the
squark and slepton masses be driven negative, and so at the intermediate
scale we generally expect the squark and slepton masses to be larger
than our previous estimate. This can be readily understood since a large
negative mass squared contribution
from a soft Higgs mass will contribute positively to the RG running
of a squark or slepton mass squared. It can also be understood from the fixed
point in Eq.\ref{XIRSFPs}. Thus the effect of the additional singlets
on the Higgs sector leads to the squark and slepton masses tending
to remain positive for longer, via the indirect effect of the Higgs masses
being driven negative more quickly, which as we have seen is not a 
problem due to the presence of the $\mu$-type parameters in the Higgs sector.

We have argued that in order to make the 
model self-consistent we should include
all the singlet couplings in the Higgs sector in our calculation.
However, as emphasised in \cite{AK}, the presence of these singlet
couplings leads to an extremely complicated system of coupled RGEs
which are difficult to solve even for the dimensionless Yukawa
couplings. The presence of all the additional soft masses will
only serve to make the calculation even more complicated, 
and we would be in danger of losing sight of the very physical
effects we are trying to illustrate. Therefore for our present purposes
we shall illustrate the effects discussed above in a simple
example which contains the essential physical features 
mentioned above. In particular our example will involve additional
Higgs Yukawa couplings, while leaving the squark and slepton RGEs
unchanged except for any changes due to the Higgs mass squared
which will feed through into these equations as discussed above.

Consider a model similar to that in Eq.\ref{W},
but augmented by three additional families of quarks and leptons,
which we distinguish by primes, which couple to the 
same Higgs as the usual three families, but which have intermediate
scale masses due to the presence of their vector partners,
i.e. we allow the same Higgs to couple to three additional families
of the form $3(\bar{5}+10)$ which have $3({5}+\bar{10})$ vector partners. 
The augmented superpotential is then: 
\begin{eqnarray}
W & = &
\sum_{i,j=1}^{3}(h_{ij}Q_{i}U_{j}^{c}H_{U_{ij}}
+k_{ij}Q_{i}D_{j}^{c}H_{D_{ij}}+l_{ij}L_{i}E_{j}^{c}H_{D_{ij}})
\nonumber \\
& + & \sum_{i,j=1}^{3}(h'_{ij}{Q'}_{i}{U'}_{j}^{c}H_{U_{ij}}
+k'_{ij}{Q'}_{i}{D'}_{j}^{c}H_{D_{ij}}+{l'}_{ij}{L'}_{i}{E'}_{j}^{c}H_{D_{ij}})
\label{Waug}
\end{eqnarray}
The merit of the couplings we have introduced is that 
they illustrate the main point concerning the 
effect of additional Higgs couplings while keeping the 
relative simplicity of the model discussed above. 
It is clear that the wavefunction anomalous dimensions of the
quark and lepton fields are unchanged from their
values in Eq.\ref{N}, and that there will be identical
quantities for the primed families with all the Yukawa couplings
being primed ones. The Higgs wavefunctions receive additional
Yukawa contributions from the primed fields:
\begin{eqnarray}
N_{H_{U_{ij}}} &=&(\frac{3}{2}r_{2}+\frac{3}{10}r_{1})\tilde{\alpha}%
_{3}-3Y^{h_{ij}}-3Y^{h'_{ij}}  \nonumber \\
N_{H_{D_{ij}}} &=&(\frac{3}{2}r_{2}+\frac{3}{10}r_{1})\tilde{\alpha}%
_{3}-3Y^{k_{ij}}-Y^{l_{ij}}-3Y^{k'_{ij}}-Y^{l'_{ij}}  \label{N'}
\end{eqnarray}

The analysis of the Yukawa fixed points goes through exactly as in
the original calculation, and we now find symmetric fixed points
with the unprimed Yukawa couplings equal to the primed Yukawa
couplings:
\begin{eqnarray}
{R^{h}}^{*} &=& {R^{h'}}^{*} =
\frac{160}{549}+\frac{21}{122}r_{2}+\frac{32}{549}r_{1}+\frac{7}{122}b_{3}
\nonumber \\
{R^{k}}^{*} &=& {R^{k}}^{*} =
\frac{176}{549}+\frac{17}{122}r_{2}-\frac{7}{2745}r_{1}+\frac{17}{366}b_{3}  
\nonumber \\
{R^{l}}^{*} &=& {R^{l}}^{*} =
-\frac{32}{183}+\frac{12}{61}r_{2}+\frac{151}{915}r_{1}+\frac{4}{61}b_{3} 
\label{FP's}
\end{eqnarray}
For example for $n=6$ we find 
${R^{h}}^{*}=0.55$, ${R^{k}}^{*}=0.52$, ${R^{l}}%
^{*}=0.145$ somewhat smaller than the values quoted in Eq.\ref{FPs}.

The RGEs for the soft scalar masses of the squarks and sleptons are 
exactly as before, and now there are additional RGEs involving the
primed squarks and sleptons which have an identical structure, but
with primes everywhere except that they involve the same (unprimed)
Higgs fields as the ordinary squarks and sleptons.
The RGEs for the soft masses of the Higgs have contributions from
both unprimed and primed squarks and sleptons which, after inserting the
Yukawa and trilinear fixed points, are:
\begin{eqnarray}
\frac{dm_{H_{U_{R_{i}}}}^{2}}{dt} &=&3(3r_{2}+\frac{3}{5}r_{1})\tilde{\alpha}%
_{3}M_{3}^{2}
-3{Y^{h}}%
^{*}(3m_{Q_{i}}^{2}+m_{U_{T}^{c}}^{2}+m_{H_{U_{R_{i}}}}^{2}+3M_{3}^{2}) 
\nonumber \\
& - & 3{Y^{h}}%
^{*}(3m_{{Q'}_{i}}^{2}+m_{{U'}_{T}^{c}}^{2}+m_{H_{U_{R_{i}}}}^{2}+3M_{3}^{2}) 
\nonumber \\
\frac{dm_{H_{U_{C_{i}}}}^{2}}{dt} &=&3(3r_{2}+\frac{3}{5}r_{1})\tilde{\alpha}%
_{3}M_{3}^{2}-3{Y^{h}}%
^{*}(m_{Q_{T}}^{2}+3m_{U_{i}^{c}}^{2}+m_{H_{U_{C_{i}}}}^{2}+3M_{3}^{2}) 
\nonumber \\
& - & 3{Y^{h}}%
^{*}(m_{{Q'}_{T}}^{2}+3m_{{U'}_{i}^{c}}^{2}+m_{H_{U_{C_{i}}}}^{2}+3M_{3}^{2}) 
\nonumber \\
\frac{dm_{H_{D_{R_{i}}}}^{2}}{dt} &=&3(3r_{2}+\frac{3}{5}r_{1})\tilde{\alpha}%
_{3}M_{3}^{2}-3{Y^{k}}%
^{*}(3m_{Q_{i}}^{2}+m_{D_{T}^{c}}^{2}+m_{H_{D_{R_{i}}}}^{2}+3M_{3}^{2}) 
\nonumber \\
&-&{Y^{l}}%
^{*}(3m_{L_{i}}^{2}+m_{E_{T}^{c}}^{2}+m_{H_{D_{R_{i}}}}^{2}+3M_{3}^{2}) 
\nonumber \\
& - & 3{Y^{k}}%
^{*}(3m_{{Q'}_{i}}^{2}+m_{{D'}_{T}^{c}}^{2}+m_{H_{D_{R_{i}}}}^{2}+3M_{3}^{2}) 
\nonumber \\
&-&{Y^{l}}%
^{*}(3m_{{L'}_{i}}^{2}+m_{{E'}_{T}^{c}}^{2}+m_{H_{D_{R_{i}}}}^{2}+3M_{3}^{2}) 
\nonumber \\
\frac{dm_{H_{D_{C_{i}}}}^{2}}{dt} &=&3(3r_{2}+\frac{3}{5}r_{1})\tilde{\alpha}%
_{3}M_{3}^{2}-3{Y^{k}}%
^{*}(m_{Q_{T}}^{2}+3m_{D_{i}^{c}}^{2}+m_{H_{D_{C_{i}}}}^{2}+3M_{3}^{2}) 
\nonumber \\
&-&{Y^{l}}%
^{*}(m_{L_{T}}^{2}+3m_{E_{i}^{c}}^{2}+m_{H_{D_{C_{i}}}}^{2}+3M_{3}^{2}) 
\nonumber \\
& - & 3{Y^{k}}%
^{*}(m_{{Q'}_{T}}^{2}+3m_{{D'}_{i}^{c}}^{2}+m_{H_{D_{C_{i}}}}^{2}+3M_{3}^{2}) 
\nonumber \\
&-&{Y^{l}}%
^{*}(m_{{L'}_{T}}^{2}+3m_{{E'}_{i}^{c}}^{2}+m_{H_{D_{C_{i}}}}^{2}+3M_{3}^{2}) 
\nonumber \\
&&
\end{eqnarray}

Due to the symmetry of the augmented model, 
It is reasonable to assume that the
primed soft masses will be driven to be equal to the unprimed
soft masses. Thus the set of RGEs in Eq.\ref{mTeqs} will
apply equally for the primed and unprimed squarks and slepton
masses, and the Higgs mass equations are simply modified
by factors of two multiplying the Yukawa couplings:
\begin{eqnarray}
\frac{dm_{H_{U_T}}^2 }{dt} &=&9(3r_{2}+\frac{3}{5}r_{1})
\tilde{\alpha}_{3}M_{3}^{2}-6{Y^{h}}^{*}(\bar{m}_U^2+9M_{3}^{2}) 
\nonumber \\
\frac{dm_{H_{D_T}}^2 }{dt} &=&9(3r_{2}+\frac{3}{5}r_{1})
\tilde{\alpha}_{3}M_{3}^{2}
- 6{Y^{k}}^{*}(\bar{m}_D^2+9M_{3}^{2})  
-2{Y^{l}}^{*}(\bar{m}_E^2 +9M_{3}^{2}) 
\label{mTeqs'}
\end{eqnarray}
These factors of two will leave Eq.\ref{mapprox} unchanged
(although the quality of the approximation may be slightly different),
but Eq.\ref{zeromodes} will become:
\begin{eqnarray}
\frac{d(2m_{Q_T}^2-m_{U_T^c}^2 -m_{D_T^c}^2 )}{dt} 
&=&3(6r_2 - \frac{19}{15}r_{1})\tilde{\alpha}_{3}M_{3}^{2}
\nonumber \\
\frac{d(2m_{L_T}^2-m_{E_T^c}^2) }{dt} 
&=&3(6r_{2}-\frac{6}{5}r_{1})\tilde{\alpha}_{3}M_{3}^{2}
\nonumber \\
\frac{d(3m_{U_T^c}^2-m_{H_{U_T}}^2  ) }{dt} 
&=&9(\frac{16}{3} - 3r_{2}+\frac{7}{15}r_{1})
\tilde{\alpha}_{3}M_{3}^{2}
\nonumber \\
\frac{d(3m_{D_T^c}^2+m_{E_T^c}^2-m_{H_{D_T}}^2  ) }{dt} 
&=& 9(\frac{16}{3} - 3r_{2}+\frac{7}{15}r_{1})
\tilde{\alpha}_{3}M_{3}^{2}
\label{zeromodes'}
\end{eqnarray}
From Eqs.\ref{mapprox}, and solution to the modified
Eq.\ref{zeromodes'} we obtain new predictions
for the individual low energy $T$ masses:
\begin{eqnarray} 
m_{Q_{T}}^{2}(t) &=&c_{Q}M_{3}^{2}(0)  
+ {\displaystyle \frac {40}{61}} \,{\it m_{Q_{T}}^{2}(0) } 
- {\displaystyle \frac {10}{61}} \,{\it m_{U_{T}^{c}}^{2}(0) }
 - {\displaystyle \frac {11}{61}} \,{\it m_{D_{T}^{c}}^{2}(0) } 
\nonumber \\ & & \mbox{} 
+ {\displaystyle \frac {2}{61}} \,{\it m_{L_{T}}^{2}(0)} 
+ {\displaystyle \frac {2}{61}} \,{\it m_{E_{T}^{c}}^{2}(0) } 
- {\displaystyle \frac {10}{183}} \,{\it m_{H_{U_T}}^2(0) } 
- {\displaystyle \frac {3}{61}} \,{\it m_{H_{D_T}}^2(0) }
\nonumber \\
m_{U_{T}^{c}}^{2}(t) &=&c_{U}M_{3}^{2}(0)  
- {\displaystyle \frac {20}{61}} \,{\it m_{Q_{T}}^{2}(0) } 
+ {\displaystyle \frac {71}{122}} \,{\it m_{U_{T}^{c}}^{2}(0) } 
+ {\displaystyle \frac {11}{122}} \,{\it m_{D_{T}^{c}}^{2}(0) } 
\nonumber \\ & & \mbox{} 
- {\displaystyle \frac {1}{61}} \,{\it m_{L_{T}}^{2}(0)} 
- {\displaystyle \frac {1}{61}} \,{\it m_{E_{T}^{c}}^{2}(0) } 
- {\displaystyle \frac {17}{122}} \,{\it m_{H_{U_T}}^2(0) }
+ {\displaystyle \frac {3}{122}} \,{\it m_{H_{D_T}}^2(0) }
\nonumber \\ 
m_{D_{T}^{c}}^{2}(t) &=&c_{D}M_{3}^{2}(0)  
- {\displaystyle \frac {22}{61}} \,{\it m_{Q_{T}}^{2}(0) } 
+ {\displaystyle \frac {11}{122}} \,{\it m_{U_{T}^{c}}^{2}(0) } 
+ {\displaystyle \frac {67}{122}} \,{\it m_{D_{T}^{c}}^{2}(0) } 
\nonumber \\ & & \mbox{} 
+ {\displaystyle \frac {5}{61}} \,{\it m_{L_{T}}^{2}(0)} 
+ {\displaystyle \frac {5}{61}} \,{\it m_{E_{T}^{c}}^{2}(0) } 
+ {\displaystyle \frac {11}{366}} \,{\it m_{H_{U_T}}^2(0) }
- {\displaystyle \frac {15}{122}} \,{\it m_{H_{D_T}}^2(0) }
\nonumber \\ 
m_{L_{T}}^{2}(t) &=&c_{L}M_{3}^{2}(0)  
+ {\displaystyle \frac {6}{61}} \,{\it m_{Q_{T}}^{2}(0) } 
- {\displaystyle \frac {3}{122}} \,{\it m_{U_{T}^{c}}^{2}(0) } 
+ {\displaystyle \frac {15}{122}} \,{\it m_{D_{T}^{c}}^{2}(0) } 
\nonumber \\ & & \mbox{} 
+ {\displaystyle \frac {43}{61}} \,{\it m_{L_{T}}^{2}(0)} 
- {\displaystyle \frac {18}{61}} \,{\it m_{E_{T}^{c}}^{2}(0) } 
- {\displaystyle \frac {1}{122}} \,{\it m_{H_{U_T}}^2(0) } 
- {\displaystyle \frac {7}{122}} \,{\it m_{H_{D_T}}^2(0) }
\nonumber \\
m_{E_{T}^{c}}^{2}(t) &=&c_{E}M_{3}^{2}(0)  
+ {\displaystyle \frac {12}{61}} \,{\it m_{Q_{T}}^{2}(0) } 
- {\displaystyle \frac {3}{61}} \,{\it m_{U_{T}^{c}}^{2}(0) } 
+ {\displaystyle \frac {15}{61}} \,{\it m_{D_{T}^{c}}^{2}(0) } 
\nonumber \\ & & \mbox{} 
- {\displaystyle \frac {36}{61}} \,{\it m_{L_{T}}^{2}(0)} 
+ {\displaystyle \frac {25}{61}} \,{\it m_{E_{T}^{c}}^{2}(0) } 
- {\displaystyle \frac {1}{61}} \,{\it m_{H_{U_T}}^2(0) } 
- {\displaystyle \frac {7}{61}} \,{\it m_{H_{D_T}}^2(0) }
\nonumber \\
m_{H_{U_T}}^2(t) &=&c_{H_U}M_{3}^{2}(0)  
- {\displaystyle \frac {60}{61}} \,{\it m_{Q_{T}}^{2}(0) } 
- {\displaystyle \frac {153}{122}} \,{\it m_{U_{T}^{c}}^{2}(0) } 
+ {\displaystyle \frac {33}{122}} \,{\it m_{D_{T}^{c}}^{2}(0) } 
\nonumber \\ & & \mbox{} 
- {\displaystyle \frac {3}{61}} \,{\it m_{L_{T}}^{2}(0)} 
- {\displaystyle \frac {3}{61}} \,{\it m_{E_{T}^{c}}^{2}(0) } 
+ {\displaystyle \frac {71}{122}} \,{\it m_{H_{U_T}}^2(0) } 
+ {\displaystyle \frac {9}{122}} \,{\it m_{H_{D_T}}^2(0) }
\nonumber \\
m_{H_{D_T}}^2(t) &=&c_{H_D}M_{3}^{2}(0)
- {\displaystyle \frac {54}{61}} \,{\it m_{Q_{T}}^{2}(0) } 
+ {\displaystyle \frac {27}{122}} \,{\it m_{U_{T}^{c}}^{2}(0) }
 - {\displaystyle \frac {135}{122}} \,{\it m_{D_{T}^{c}}^{2}(0) } 
\nonumber \\ & & \mbox{} 
- {\displaystyle \frac {21}{61}} \,{\it m_{L_{T}}^{2}(0)} 
- {\displaystyle \frac {21}{61}} \,{\it m_{E_{T}^{c}}^{2}(0) } 
+ {\displaystyle \frac {9}{122}} \,{\it m_{H_{U_T}}^2(0) } 
+ {\displaystyle \frac {63}{122}} \,{\it m_{H_{D_T}}^2(0) }
\label{fullmTpredictions'}
\end{eqnarray}
As before in the gauge dominated scenario we obtain
the simple predictions
\begin{eqnarray}
m_{Q_{T}}^{2}(t) &\approx &c_{Q}M_{3}^{2}(0)  \nonumber \\
m_{U_{T}^{c}}^{2}(t) &\approx &c_{U}M_{3}^{2}(0)  \nonumber \\
m_{D_{T}^{c}}^{2}(t) &\approx &c_{D}M_{3}^{2}(0)  \nonumber \\
m_{L_{T}}^{2}(t) &\approx &c_{L}M_{3}^{2}(0)  \nonumber \\
m_{E_{T}^{c}}^{2}(t) &\approx &c_{E}M_{3}^{2}(0)  \nonumber \\
m_{H_{U_T}}^2(t) &\approx &c_{H_U}M_{3}^{2}(0)  \nonumber \\
m_{H_{D_T}}^2(t) &\approx &c_{H_D}M_{3}^{2}(0)
\label{mTpredictions2}
\end{eqnarray}
where the new $c_i$ coefficients 
appropriate to this case are given in Table 3.
Note that the squark and slepton mass squareds are always
positive, but that the Higgs mass squareds remain negative
at the intermediate scale.
Comparing the coefficients in Table 3 to those in Table 2 it is seen that
the effect of the additional Higgs couplings is to drive the Higgs mass
squareds more negative, and hence keep the squark and slepton mass
squareds positive at the intermediate scale, as argued previously.
One should not be alarmed that the Higgs mass squareds are negative 
at the intermediate scale. Indeed in the MSSM it is commonplace for 
Higgs mass squareds to be driven negative one or two orders of magnitude
below the unification scale. What enters the effective Higgs potential
is the combination $\mu^2 + m_H^2$, so a suitable
choice of $\mu$ parameter will give correct low energy electroweak
symmetry breaking as in the MSSM. We will give a detailed discussion
of this in section 7.

\begin{table}[tbp]
\hfil
\begin{tabular}{cccccccc}
\hline
$n$ & $c_Q$ & $c_U$ & $c_D$ & $c_L$ & $c_E$ & $c_{H_U}$ & $c_{H_D}$\\ \hline
6 & 1.04 & 0.52 & 0.42 & 0.87 & 0.59 & -4.67 & -4.38 \\ \hline
8 & 0.66 & 0.24 & 0.20 & 0.58 & 0.28 & -2.71 & -2.57 \\ \hline
10 & 0.48 & 0.15 & 0.12 & 0.44 & 0.17 & -1.89 & -1.81  \\ \hline
20 & 0.21 & 0.036 & 0.029 & 0.20 & 0.040 & -0.74 & -0.72 \\ \hline
40 & 0.099 & 0.011 & 0.009 & 0.096 & 0.012 & -0.33 & -0.33 \\ \hline
\end{tabular}
\hfil
\caption{$c_Q,c_U,c_D,c_L,c_E,c_{H_U},c_{H_D}$ 
coefficients as a function of $n$ for the augmented model
with additional Higgs couplings.}
\end{table}

\section{Infra-red structure for the family differences of soft scalar masses}

Since the fixed point structure makes definite predictions the soft squark and 
slepton masses,
c.f. Table 3, we are in a position to ask whether it is capable of giving an 
explanation of the flavour problem in supersymmetry. This requires that the
squark and slepton mass splittings between different families
be small compared to the squark and slepton masses themselves,
in order to suppress flavour changing neutral currents. We will illustrate the 
structure by considering  Eq.\ref{mIRSFPRGEs} for the original formulation of 
the model without
extra Higgs couplings. 
Although this model is not viable for the reasons 
discussed it contains the general flavour structure 
in a simple form and is thus a useful starting point. 
We will return to the more realistic models later. 

Consider the following family differences in soft scalar masses :
\begin{eqnarray}
\Delta _{Q_{ij}}^{2} &\equiv &m_{Q_{i}}^{2}-m_{Q_{j}}^{2}  \nonumber \\
\Delta _{U_{ij}^{c}}^{2} &\equiv &m_{U_{i}^{c}}^{2}-m_{U_{j}^{c}}^{2} 
\nonumber \\
\Delta _{D_{ij}^{c}}^{2} &\equiv &m_{D_{i}^{c}}^{2}-m_{D_{j}^{c}}^{2} 
\nonumber \\
\Delta _{L_{ij}}^{2} &\equiv &m_{L_{i}}^{2}-m_{L_{j}}^{2}  \nonumber \\
\Delta _{E_{ij}^{c}}^{2} &\equiv &m_{E_{i}^{c}}^{2}-m_{E_{j}^{c}}^{2} 
\nonumber \\
\Delta _{H_{U_{R_{ij}}}}^{2} &\equiv
&m_{H_{U_{R_{i}}}}^{2}-m_{H_{U_{R_{j}}}}^{2}  \nonumber \\
\Delta _{H_{U_{C_{ij}}}}^{2} &\equiv
&m_{H_{U_{C_{i}}}}^{2}-m_{H_{U_{C_{j}}}}^{2}  \nonumber \\
\Delta _{H_{D_{R_{ij}}}}^{2} &\equiv
&m_{H_{D_{R_{i}}}}^{2}-m_{H_{D_{R_{j}}}}^{2}  \nonumber \\
\Delta _{H_{D_{C_{ij}}}}^{2} &\equiv
&m_{H_{D_{C_{i}}}}^{2}-m_{H_{D_{C_{j}}}}^{2}  \label{Delta}
\end{eqnarray}

We now return to Eq.\ref{rge1} and
we construct the RGEs for the differences $\Delta _{ij}^{2}$.
The gauge parts of the RGEs cancel
and the structure of the RGEs becomes independent
of the choice of flavour indices $i,j$:
\begin{eqnarray}
\frac{d\Delta _{U_{ij}^{c}}^{2}}{dt} & = & -2{Y^{h}}^{*}
{\bar{\Delta}_{U_{ij}^{c}}^{2}} \nonumber \\
\frac{d{\Delta_{H_{U_{C_{ij}}}}^{2}}}{dt} & = & -3{Y^{h}}^{*}
{\bar{\Delta}_{U_{ij}^{c}}^{2}} \nonumber \\
\frac{d\Delta _{D_{ij}^{c}}^{2}}{dt} & = & -2{Y^{k}}^{*}
{\bar{\Delta}_{D_{ij}^{c}}^{2}} \nonumber \\
\frac{d\Delta _{E_{ij}^{c}}^{2}}{dt} & = & -2{Y^{l}}^{*}
{\bar{\Delta}_{E_{ij}^{c}}^{2}} \nonumber \\
\frac{d{\Delta_{H_{D_{C_{ij}}}}^{2}}}{dt} & = & -3{Y^{k}}^{*}
{\bar{\Delta}_{D_{ij}^{c}}^{2}} -{Y^{l}}^{*}
{\bar{\Delta}_{E_{ij}^{c}}^{2}} \nonumber \\
\frac{d\Delta _{Q_{ij}}^{2}}{dt} & = & -{Y^{h}}^{*}
\bar{\Delta}_{{QU}_{ij}}^{2} -{Y^{k}}^{*}
\bar{\Delta}_{{QD}_{ij}}^{2} 
\nonumber \\
\frac{d\Delta _{L_{ij}}^{2}}{dt} & = & -{Y^{l}}^{*}
\bar{\Delta}_{{L}_{ij}}^{2} 
\nonumber \\
\frac{d{\Delta_{H_{U_{R_{ij}}}}^{2}}}{dt} & = & -3{Y^{h}}^{*}
\bar{\Delta}_{{QU}_{ij}}^{2} 
\nonumber \\
\frac{d{\Delta_{H_{D_{R_{ij}}}}^{2}}}{dt} & = & -3{Y^{k}}^{*}
\bar{\Delta}_{{QD}_{ij}}^{2}-{Y^{l}}^{*}
\bar{\Delta}_{{L}_{ij}}^{2}  
\label{Deltaeqs}
\end{eqnarray}
where we have defined the following combinations:
\begin{eqnarray}
{\bar{\Delta}_{U_{ij}^{c}}^{2}} & = &
3{\Delta _{U_{ij}^{c}}^{2}}+{\Delta_{H_{U_{C_{ij}}}}^{2}}  \nonumber \\
{\bar{\Delta}_{D_{ij}^{c}}^{2}} & = &
3{\Delta _{D_{ij}^{c}}^{2}}+{\Delta_{H_{D_{C_{ij}}}}^{2}}  \nonumber \\
{\bar{\Delta}_{E_{ij}^{c}}^{2}} & = &
3{\Delta _{E_{ij}^{c}}^{2}}+{\Delta_{H_{D_{C_{ij}}}}^{2}}  \nonumber \\
\bar{\Delta}_{{QU}_{ij}}^{2} & = &
3\Delta _{Q_{ij}}^{2}+{\Delta_{H_{U_{R_{ij}}}}^{2}} \nonumber \\
\bar{\Delta}_{{QD}_{ij}}^{2} & = &
3\Delta _{Q_{ij}}^{2}+{\Delta_{H_{D_{R_{ij}}}}^{2}} \nonumber \\
\bar{\Delta}_{{L}_{ij}}^{2} & = &
3\Delta _{L_{ij}}^{2}+{\Delta_{H_{D_{R_{ij}}}}^{2}}
\label{Deltabar}
\end{eqnarray}
Notice that the RGEs decouple into three
different sectors: 
\begin{eqnarray}
&&[{\Delta _{U_{ij}^{c}}^{2}},{\Delta_{H_{U_{C_{ij}}}}^{2}}],  \nonumber \\
&&[{\Delta _{D_{ij}^{c}}^{2}},{\Delta _{E_{ij}^{c}}^{2}},
{\Delta _{H_{D_{C_{ij}}}}^{2}}],  \nonumber \\
&&[\Delta _{Q_{ij}}^{2},\Delta _{L_{ij}}^{2},
{\Delta_{H_{U_{R_{ij}}}}^{2}},{\Delta _{H_{D_{R_{ij}}}}^{2}}]
\label{Deltasectors}
\end{eqnarray}
Also note that the 9 RGEs only depend on 6 independent combinations of the
$\Delta$'s which satisfy the RGEs:
\begin{eqnarray}
\frac{d\bar{\Delta}_{U_{ij}^{c}}^{2}}{dt} & = & -9{Y^{h}}^{*}
{\bar{\Delta}_{U_{ij}^{c}}^{2}} \nonumber \\
\frac{d\bar{\Delta}_{D_{ij}^{c}}^{2}}{dt} & = & -9{Y^{k}}^{*}
{\bar{\Delta}_{D_{ij}^{c}}^{2}} -{Y^{l}}^{*}
{\bar{\Delta}_{E_{ij}^{c}}^{2}} \nonumber \\
\frac{d\bar{\Delta}_{E_{ij}^{c}}^{2}}{dt} & = & -3{Y^{k}}^{*}
{\bar{\Delta}_{D_{ij}^{c}}^{2}} -7{Y^{l}}^{*}
{\bar{\Delta}_{E_{ij}^{c}}^{2}} \nonumber \\
\frac{d\bar{\Delta}_{{QU}_{ij}}^{2} }{dt} & = & -6{Y^{h}}^{*}
\bar{\Delta}_{{QU}_{ij}}^{2}     -3{Y^{k}}^{*}
\bar{\Delta}_{{QD}_{ij}}^{2} \nonumber \\
\frac{d\bar{\Delta}_{{QD}_{ij}}^{2} }{dt} & = & -3{Y^{h}}^{*}
\bar{\Delta}_{{QU}_{ij}}^{2}     -6{Y^{k}}^{*}
\bar{\Delta}_{{QD}_{ij}}^{2} -{Y^{l}}^{*}
\bar{\Delta}_{{L}_{ij}}^{2} 
\nonumber \\
\frac{d\bar{\Delta}_{{L}_{ij}}^{2} }{dt} & = & -3{Y^{k}}^{*}
\bar{\Delta}_{{QD}_{ij}}^{2} -4{Y^{l}}^{*}
\bar{\Delta}_{{L}_{ij}}^{2} 
\label{Deltabareqs}
\end{eqnarray}
The first of these equations is already diagonal, but 
the other two sectors involve coupled equations which must be diagonalised.
After
diagonalisation of each of the three sectors the equations take the form: 
\begin{equation}
\frac{d\bar{\Delta}_{i}^{2}}{dt}=-C_{i}\tilde{\alpha}_{3}
\bar{\Delta}_{i}^{2}
\label{DeltaRGE}
\end{equation}
where $i=1$ for the first sector, $i=2,3$ for
the second sector and $i=4,\cdots 6$ for the third sector and the 
$C_{i}$ are shown in Table 4.
The equations have solution: 
\begin{equation}
\bar{\Delta}_{i}^{2}(t)
=\bar{\Delta}_{i}^{2}(0)\left( \frac{\tilde{\alpha}_{3}(t)}{%
\tilde{\alpha}_{3}(0)}\right) ^{\frac{C_{i}}{b_{3}}}  \label{Deltasoln1}
\end{equation}

\begin{table}[tbp]
\hfil
\begin{tabular}{ccccccc}
\hline
$n$ & $C_1$ & $C_2$ & $C_3$ & $C_4$ & $C_5$ & $C_6$
\\ \hline
6 & 6.0 & 5.7 & 1.85 & 5.8 & 2.2 & 0.83 \\ \hline
8 & 7.5 & 7.2 & 3.2  & 7.3 & 3.0 & 1.3 \\ \hline
10 & 8.9 & 8.6 & 4.4 & 8.7 & 3.7 & 1.7 \\ \hline
20 & 15.6 & 15.4 & 9.5 & 15.0 & 7.5 & 3.36 \\ \hline
40 & 28.2 & 29.1 & 18.6 & 27.3 & 14.7 & 6.3 \\ \hline
\end{tabular}
\hfil
\caption{$C_i$ coefficients as a function of $n$.}
\end{table}

The orthogonal combinations of soft scalar mass differences, again not unique
due to the degeneracy, have zero beta functions. A simple choice is:
\begin{eqnarray}
\frac{d(3\Delta_{U_{ij}^{c}}^{2}-2\Delta_{H_{U_{C_{ij}}}}^{2} )}{dt} & = & 0
\nonumber \\
\frac{d(3\Delta_{D_{ij}^{c}}^{2}+\Delta_{E_{ij}^{c}}^{2}
-2\Delta_{H_{D_{C_{ij}}}}^{2})}{dt} & = & 0
\nonumber \\
\frac{d(3\Delta_{Q_{ij}}^{2}+\Delta_{L_{ij}}^{2}
-\Delta_{H_{U_{R_{ij}}}}^{2}-\Delta_{H_{D_{R_{ij}}}}^{2})}{dt} & = & 0
\label{Deltazeroeqs}
\end{eqnarray}
Clearly Eq.\ref{Deltazeroeqs} has the trivial solutions:
\begin{eqnarray}
3\Delta_{U_{ij}^{c}}^{2}(t)-2\Delta_{H_{U_{C_{ij}}}}^{2}(t)  & = & 
3\Delta_{U_{ij}^{c}}^{2}(0)-2\Delta_{H_{U_{C_{ij}}}}^{2}(0)
\nonumber \\
3\Delta_{D_{ij}^{c}}^{2}(t)+\Delta_{E_{ij}^{c}}^{2}(t)
-2\Delta_{H_{D_{C_{ij}}}}^{2}(t) & = &  
3\Delta_{D_{ij}^{c}}^{2}(0)+\Delta_{E_{ij}^{c}}^{2}(0)
-2\Delta_{H_{D_{C_{ij}}}}^{2}(0)
\nonumber \\
3\Delta_{Q_{ij}}^{2}(t)+\Delta_{L_{ij}}^{2}(t)
-\Delta_{H_{U_{R_{ij}}}}^{2}(t)-\Delta_{H_{D_{R_{ij}}}}^{2}(t) & = &
3\Delta_{Q_{ij}}^{2}(0)+\Delta_{L_{ij}}^{2}(0)
\nonumber \\  
& - & \Delta_{H_{U_{R_{ij}}}}^{2}(0)-\Delta_{H_{D_{R_{ij}}}}^{2}(0)
\label{Deltazeroslns}
\end{eqnarray}

Clearly these $\Delta$ combinations 
corresponding to zero beta functions will have low
energy values equal to their high energy boundary values at $t=0$, while the 
$\bar{\Delta}_{i}^{2}$ 
corresponding to non-zero eigenvalues will have values at
low energy suppressed by the anomalous dimension according to the values of $%
C_{i}$ in Table 4. The suppression factor coming from Eq.\ref{Deltasoln1}
is $\left( \frac{\tilde{\alpha}_{3}(t)}{%
\tilde{\alpha}_{3}(0)}\right) ^{\frac{C_{i}}{b_{3}}}$.
According to Table 4 ${\frac{C_{i}}{n-3}}>1$ for $C_1,C_2,C_4$,
but ${\frac{C_{i}}{n-3}}<1$ for $C_3,C_5$ and in particular
for $C_6$ the anomalous dimension is significantly smaller than unity.
Nevertheless as a rough approximation we shall
assume $\bar{\Delta}_i(t)\approx 0$ for all these quantities
so that from Eq.\ref{Deltazeroslns}
we find the following infra red structure which applies at the intermediate 
scale : 
\begin{eqnarray}
\Delta _{U_{ij}^{c}}^{2}(t) &\approx &
\frac{1}{3}\Delta _{U_{ij}^{c}}^{2}(0)-\frac{2%
}{9}\Delta _{H_{U_{C_{ij}}}}^{2}(0)  \nonumber \\
\Delta _{D_{ij}^{c}}^{2}(t) &\approx &
\frac{3}{10}\Delta _{D_{ij}^{c}}^{2}(0)+\frac{%
1}{10}\Delta _{E_{ij}^{c}}^{2}(0)-\frac{2}{10}\Delta _{H_{D_{C_{ij}}}}^{2}(0)
\nonumber \\
\Delta _{E_{ij}^{c}}^{2}(t) &\approx &
\frac{3}{10}\Delta _{D_{ij}^{c}}^{2}(0)+\frac{%
1}{10}\Delta _{E_{ij}^{c}}^{2}(0)-\frac{2}{10}\Delta _{H_{D_{C_{ij}}}}^{2}(0)
\nonumber \\
\Delta _{Q_{ij}}^{2}(t) &\approx &
\frac{3}{10}\Delta _{Q_{ij}}^{2}(0)+\frac{1}{10}%
\Delta _{L_{ij}}^{2}(0)-\frac{1}{10}\Delta _{H_{U_{R_{ij}}}}^{2}(0)-\frac{1}{%
10}\Delta _{H_{D_{R_{ij}}}}^{2}(0)  \nonumber \\
\Delta _{L_{ij}}^{2}(t) &\approx &
\frac{3}{10}\Delta _{Q_{ij}}^{2}(0)+\frac{1}{10}%
\Delta _{L_{ij}}^{2}(0)-\frac{1}{10}\Delta _{H_{U_{R_{ij}}}}^{2}(0)-\frac{1}{%
10}\Delta _{H_{D_{R_{ij}}}}^{2}(0)  \nonumber \\
\Delta _{H_{U_{R_{ij}}}}^{2}(t) &\approx &
-\frac{9}{10}\Delta _{Q_{ij}}^{2}(0)-%
\frac{3}{10}\Delta _{L_{ij}}^{2}(0)+\frac{3}{10}\Delta
_{H_{U_{R_{ij}}}}^{2}(0)+\frac{3}{10}\Delta _{H_{D_{R_{ij}}}}^{2}(0) 
\nonumber \\
\Delta _{H_{D_{R_{ij}}}}^{2}(t) &\approx &
-\frac{9}{10}\Delta _{Q_{ij}}^{2}(0)-%
\frac{3}{10}\Delta _{L_{ij}}^{2}(0)+\frac{3}{10}\Delta
_{H_{U_{R_{ij}}}}^{2}(0)+\frac{3}{10}\Delta _{H_{D_{R_{ij}}}}^{2}(0)
\nonumber \\
\Delta _{H_{U_{C_{ij}}}}^{2}(t) & \approx &
-\Delta _{U_{ij}^{c}}^{2}(0)+\frac{2}{3}%
\Delta _{H_{U_{C_{ij}}}}^{2}(0)  
\nonumber \\
\Delta _{H_{D_{C_{ij}}}}^{2}(t) &\approx &
-\frac{9}{10}\Delta _{D_{ij}^{c}}^{2}(0)-%
\frac{3}{10}\Delta _{E_{ij}^{c}}^{2}(0)+\frac{6}{10}\Delta
_{H_{D_{C_{ij}}}}^{2}(0)  
\label{Deltasoln2}
\end{eqnarray}
Note that the infra red structure leads to the phenomenologically interesting 
predictions 
\begin{equation}
\Delta _{D_{ij}^{c}}^{2}(t)\approx\Delta _{E_{ij}^{c}}^{2}(t), \  \ 
\Delta _{Q_{ij}}^{2}(t)\approx \Delta _{L_{ij}}^{2}(t)
\end{equation}
valid at the intermediate scale, will for the light generations also apply at 
low energy scales.  

The results in Eq.\ref{Deltasoln2}, which follow from Eq.\ref{Deltazeroslns}
together with the approximation that all the
$\bar{\Delta}_i^2(t)\approx 0$ due to anomalous dimension suppression,
are at first sight rather disappointing 
for they fall short of a complete 
solution to the flavour problem. 
The reason may be seen from Eq.\ref{Deltazeroeqs}. The fact that 
these combinations of soft scalar masses 
do not evolve means initial flavour 
non-degeneracy in these channels 
will not be erased. 
This effect is quite general and follows 
because, c.f. Eq.\ref{rge1}, 
the right hand side of the renormalisation group 
equations depend on the soft scalar masses through a reduced 
number of terms involving the sums of the soft scalar masses 
of the states involved in the Yukawa coupling responsible 
for the particular term. As a result it is always possible 
to eliminate the term via a combination of the RGE involving 
the same Yukawa coupling. This is the reason that infra-red 
structure cannot completely solve the flavour problem.

The effects of partial Higgs cancellation, and numerical
suppression of the flavour differences observed in the solutions
can be understood by returning to the original
RGEs for the $\Delta$'s in Eq.\ref{Deltaeqs} where it is observed that
the $\Delta$'s corresponding to the Higgs are driven to be smaller
faster than for the squarks and sleptons. Thus there will be a region
of running where the Higgs $\Delta$'s are positive but small
compared to the squark and slepton $\Delta$'s, and so may be neglected
over this region. Over the region where the 
Higgs $\Delta$'s are negligible then
it is clear from Eq.\ref{Deltaeqs} that all the squark and slepton
masses will be suppressed by anomalous dimensions since the
$\bar{\Delta}$'s become equal to the $\Delta$'s (up to a factor of 3),
and there are no channels with zero beta function.
For example in the extreme limit that all the Higgs mass squared
differences were negligible over the entire range
the effect of the fixed point would be to suppress all of the
squark and slepton flavour mass squared differences by powers
of the anomalous dimension. In reality the Higgs $\Delta$'s 
are non-negligible to begin with and also may be
driven to be large and negative,
so it is not a good appproximation to neglect them from the
RGEs of the squark and slepton mass squared differences.
However the effect the Higgs $\Delta$'s being driven smaller
more quickly than the squark and slepton $\Delta$'s
is imprinted onto the final results in Eq.\ref{Deltasoln2}, and accounts
for the suppressions of flavour differences discussed above.

Similar cancellation effects also occur in the family sums over soft scalar
masses,
so it will not lead to a significant suppression of flavour changing
neutral currents which depend on ratios like  $\Delta_{ij}^2 /m_T^2$.
An efficient way to suppress flavour changing effects in this model is
to assume gaugino dominance near the string scale. 
As discussed in the previous section, gaugino dominance 
is the statement that the soft scalar masses at high energy (string) scale
are negligible compared to the high energy gaugino masses,
$m_i^2(0)\ll M_3^2(0)$. As regards the mass differences here,
it is clear that the RGEs preserve the flavour differences
at the string scale down to the intermediate scale.
Clearly then if the actual magnitudes of the soft scalar
masses near the string scale
are very small then their differences must also be small, and this smallness
will be preserved by the fixed point structure. By contrast we
showed that the low energy (intermediate
scale) family sums of soft scalar masses were given by Eq.\ref{mTpredictions}, 
and are of order the gaugino mass near the string scale
$M_3^2(0)$, which is much larger than the soft scalar
masses near the string scale
by assumption. Therefore the assumption of gaugino dominance solves
the flavour problem in this kind of theory. 
Of course below the intermediate scale the third family Yukawa couplings
will dominate and induce flavour violations at low energy, but it is
important to note that in this theory this effect does not begin until
below the intermediate scale, rather than below the string scale as in the
MSSM, so the effects of flavour violation will be smaller in this case.
Moreover, as we shall see in the next section, the spectrum which
results from the assumption of gaugino dominance is more acceptable
phenomenologically than in the MSSM.
For these reasons, we find gaugino dominance a very appealing possibility
in this model.

To end this section we remark that in the augmented model with extra
Higgs couplings the qualitative picture for the $\Delta$'s
does not change very much. Assuming symmetry between the primed and
unprimed masses, all that happens is that the RGEs for the
Higgs $\Delta$'s all have their beta functions multiplied by
an overall factor of 2, so that they get driven smaller more quickly.
This means that for example we find
\begin{eqnarray}
\frac{d(6\Delta_{U_{ij}^{c}}^{2}-2\Delta_{H_{U_{C_{ij}}}}^{2} )}{dt} & = & 0
\nonumber \\
\frac{d(6\Delta_{D_{ij}^{c}}^{2}+2\Delta_{E_{ij}^{c}}^{2}
-2\Delta_{H_{D_{C_{ij}}}}^{2})}{dt} & = & 0
\nonumber \\
\frac{d(6\Delta_{Q_{ij}}^{2}+2\Delta_{L_{ij}}^{2}
-\Delta_{H_{U_{R_{ij}}}}^{2}-\Delta_{H_{D_{R_{ij}}}}^{2})}{dt} & = & 0
\label{Deltazeroeqs'}
\end{eqnarray}
in place of Eq.\ref{Deltazeroeqs}, and so on.
The fact that the Higgs mass squared differences get driven smaller
more quickly in this case will lead to even greater suppression
of the low energy squark and slepton mass squared flavour differences
than in the original model. However precise predictions are 
more difficult in this case due to the need to make some assumption about the 
relative magnitude of
the primed and unprimed soft masses. 
If the assumption of equal soft masses in the primed and unprimed
sectors is relaxed then this also does not qualitatively change the
conclusions since we find:
\begin{eqnarray}
\frac{d(3\Delta_{U_{ij}^{c}}^{2}
+3\Delta_{{U'}_{ij}^{c}}^{2}
-2\Delta_{H_{U_{C_{ij}}}}^{2} )}{dt} & = & 0
\nonumber \\
\frac{d(3\Delta_{D_{ij}^{c}}^{2}
+3\Delta_{{D'}_{ij}^{c}}^{2}
+\Delta_{E_{ij}^{c}}^{2}+2\Delta_{{E'}_{ij}^{c}}^{2}
-2\Delta_{H_{D_{C_{ij}}}}^{2})}{dt} & = & 0
\nonumber \\
\frac{d(3\Delta_{Q_{ij}}^{2}+3\Delta_{{Q'}_{ij}}^{2}
+\Delta_{L_{ij}}^{2}+\Delta_{{L'}_{ij}}^{2}
-\Delta_{H_{U_{R_{ij}}}}^{2}-\Delta_{H_{D_{R_{ij}}}}^{2})}{dt} & = & 0
\label{Deltazeroeqs''}
\end{eqnarray}
Again we see the appearance of combinations
with zero beta functions, leading to qualitatively similar results
to those obtained previously.

\section{The Low Energy Spectrum}

We have already remarked that the assumption of
gaugino dominance leads to an inverted spectrum in strong unification
as compared to the MSSM. For this reason it is worthwhile to investigate
the typical features of the spectrum in this case.
In order to investigate the low energy spectrum we need to use
the strong unification predictions at the intermediate scale
as boundary conditions for the MSSM which is the effective
low energy theory below this scale. There are various possibilities 
depending on the make-up of the two light Higgs doublets of the MSSM 
in terms of the 
9 Higgs doublets coupling to the up-sector, $H_U$, and the 9 Higgs doublets
coupling to the down- and charged lepton-sector, $H_D$.

The first possibility is the conventional one in which the two Higgs doublets 
are closely identified with
the $H_{U_{33}}$ and $H_{D_{33}}$ doublets of the high energy theory.
This gives large, almost equal, values of the third family MSSM
Yukawa couplings as prescribed by the fixed point, corresponding
to a large value of $\tan \beta \approx m_t/m_b$ 
to account for the top-bottom mass difference.
A second possibility is that 
$H_U\approx H_{U_{33}}$ but $H_D\approx H_X+\epsilon H_{D_{33}}$ 
where $H_X$ is a Higgs doublet which does not couple to quarks and 
leptons and $\epsilon \sim m_b/m_t$ is a small parameter
\footnote{Such a scheme may arises naturally in string compactification with, 
for example, $H_X=\bar{H}_{U_{33}}$. 
It often happens that the barred fields correspond to (1,1) 
forms while the un-barred fields correspond to (2,1) forms. 
Trilinear couplings do not then mix barred and unbarred fields.}. 
This effectively reduces the MSSM bottom and tau Yukawa couplings at the 
intermediate scale, and so allows values of $\tan \beta \sim 1$.
A third possibility has the two Higgs doublets
as democratic mixtures of all the Higgs doublets,
$H_U = \frac{1}{3}\sum H_{U_{ij}}$, and $H_D = \frac{1}{3}\sum H_{D_{ij}}$.
In this case the MSSM can only be rescued by major surgery \cite{Abel}.

We shall concentrate in this section on only the 
simplest first case corresponding to large and approximately equal
third family Yukawa couplings. Of course the disadvantage of this is that
there will be the usual fine-tuning required in order to
sustain the large hierarchy of vacuum expectation values. 
However we shall not worry about such matters
here. Including the third family Yukawa couplings, the RGEs for the
third family in the MSSM correspond to taking the 33
components of the RGEs given in section 2
with $Y^{h_{33}}\equiv Y_t$, $A^{h_{33}}\equiv A_t$,
$m_{H_{U_{33}}}^2\equiv m_{H_U}^2$ etc. 
Although these MSSM RGEs
are well known we reproduce them here for purposes of comparison
to the RGEs in section 2. Analagous to the Yukawa RGEs in Eq.\ref{YukRGEs}
we have:
\begin{eqnarray}
\frac{dY_t}{dt} &=&Y_t(\frac{16}{3}\tilde{\alpha}_3 
+ 3\tilde{\alpha}_2 + \frac{13}{15} \tilde{\alpha}_1-6Y_t-Y_b)
\nonumber \\
\frac{dY_b}{dt} &=&Y_b(\frac{16}{3}\tilde{\alpha}_3 
+ 3\tilde{\alpha}_2 + \frac{7}{15} \tilde{\alpha}_1-6Y_b-Y_t-Y_{\tau})
\nonumber \\
\frac{dY_{\tau}}{dt} &=&Y_{\tau}(
3\tilde{\alpha}_2 + \frac{9}{5} \tilde{\alpha}_1-4Y_{\tau}-3Y_b)
\nonumber \\
\label{YukMSSM}
\end{eqnarray}

The trilinear RGEs analagous to Eq.\ref{TrilinearRGEs} are:
\begin{eqnarray}
\frac{dA_t}{dt} &=&\frac{16}{3}\tilde{\alpha}_3M_3 
+ 3\tilde{\alpha}_2M_2 + \frac{13}{15}\tilde{\alpha}_1-6A_tY_t-A_bY_b
\nonumber \\
\frac{dA_b}{dt} &=&\frac{16}{3}\tilde{\alpha}_3M_3 
+ 3\tilde{\alpha}_2M_2 + \frac{7}{15}\tilde{\alpha}_1M_1
-6A_bY_b-A_tY_t-A_{\tau}Y_{\tau}
\nonumber \\
\frac{dA_{\tau}}{dt} &=&3\tilde{\alpha}_2M_2 + \frac{9}{5}\tilde{\alpha}_1M_1
-4A_{\tau}Y_{\tau}-3A_bY_b
\nonumber \\
\label{AMSSM}
\end{eqnarray}

The soft scalar mass RGEs analagous to Eq.\ref{mRGEs} are:
\begin{eqnarray}
\frac{dm_{Q_3}^2}{dt} &=&
\frac{16}{3}\tilde{\alpha}_3 M_3^2
 + 3\tilde{\alpha}_2 M_2^2 + \frac{1}{15}\tilde{\alpha}_1 M_1^2
-Y_t(\bar{m}_{U_3}^2 +A_t^2)  - Y_b(\bar{m}_{D_3}^2+A_b^2) 
\nonumber \\
\frac{dm_{U_3^c}^2}{dt} &=&
\frac{16}{3}\tilde{\alpha}_3 M_3^2
+ \frac{16}{15}\tilde{\alpha}_1 M_1^2
-2Y_t(\bar{m}_{U_3}^2 +A_t^2)
\nonumber \\
\frac{dm_{D_3^c}^2 }{dt} &=&
\frac{16}{3}\tilde{\alpha}_3 M_3^2
+ \frac{4}{15}\tilde{\alpha}_1 M_1^2
-2Y_b(\bar{m}_{D_3}^2 +A_b^2)
\nonumber \\
\frac{dm_{L_3}^2 }{dt} &=&
3\tilde{\alpha}_2 M_2^2 + \frac{3}{5}\tilde{\alpha}_1 M_1^2
-Y_{\tau}(\bar{m}_{E_3}^2 +A_{\tau}^2)
\nonumber \\
\frac{dm_{E_3^c}^2 }{dt} &=&
\frac{12}{5}\tilde{\alpha}_1 M_1^2
-2Y_{\tau}(\bar{m}_{E_3}^2 +A_{\tau}^2)
\nonumber \\
\frac{dm_{H_{U}}^2 }{dt} &=&
3\tilde{\alpha}_2 M_2^2 + \frac{9}{15}\tilde{\alpha}_1 M_1^2
-3Y_t(\bar{m}_{U_3}^2 +A_t^2)
\nonumber \\
\frac{dm_{H_{D}}^2 }{dt} &=&
3\tilde{\alpha}_2 M_2^2 + \frac{3}{5}\tilde{\alpha}_1 M_1^2
-3Y_b(\bar{m}_{D_3}^2 +A_b^2)
-Y_{\tau}(\bar{m}_{E_3}^2 +A_{\tau}^2)
\label{mMSSM}
\end{eqnarray}
where we have defined the combinations analagous to those in Eq.\ref{mbar}:
\begin{eqnarray}
\bar{m}_{U_3}^2 & = & m_{Q_3}^2+m_{U_3^c}^2+m_{H_{U}}^2 \nonumber \\
\bar{m}_{D_3}^2 & = & m_{Q_3}^2+m_{D_3^c}^2+m_{H_{D}}^2 \nonumber \\
\bar{m}_{E_3}^2 & = & m_{L_3}^2+m_{E_3^c}^2+m_{H_{D}}^2
\label{mbarMSSMdefn}
\end{eqnarray}
These combinations satisfy RGEs analagous to Eq.\ref{nonzeromodes}: 
\begin{eqnarray}
\frac{d\bar{m}_{U_3}^2}{dt} &=&
\frac{32}{3}\tilde{\alpha}_3 M_3^2
 + 6\tilde{\alpha}_2 M_2^2 + \frac{26}{15}\tilde{\alpha}_1 M_1^2
-6Y_t(\bar{m}_{U_3}^2 +A_t^2)  - Y_b(\bar{m}_{D_3}^2+A_b^2) 
\nonumber \\
\frac{d\bar{m}_{D_3}^2}{dt} &=&
\frac{32}{3}\tilde{\alpha}_3 M_3^2
+ 6\tilde{\alpha}_2 M_2^2 + \frac{14}{15}\tilde{\alpha}_1 M_1^2
-6Y_b(\bar{m}_{D_3}^2+A_b^2)-Y_t(\bar{m}_{U_3}^2 +A_t^2)  
-Y_{\tau}(\bar{m}_{E_3}^2 +A_{\tau}^2)
\nonumber \\
\frac{d\bar{m}_{E_3}^2}{dt} &=&
6\tilde{\alpha}_2 M_2^2
+ \frac{18}{5}\tilde{\alpha}_1 M_1^2
-4Y_{\tau}(\bar{m}_{E_3}^2 +A_{\tau}^2)-3Y_b(\bar{m}_{D_3}^2 +A_b^2)
\label{nonzeromodesMSSM}
\end{eqnarray}
The RGEs for the orthogonal 
combinations analagous to Eq.\ref{zeromodes} are:
\begin{eqnarray}
\frac{d(2m_{Q_3}^2-m_{U_3^c}^2 -m_{D_3^c}^2 )}{dt} 
&=&6\tilde{\alpha}_{2}M_{2}^{2} 
- \frac{18}{15}\tilde{\alpha}_{1}M_{1}^{2}
\nonumber \\
\frac{d(2m_{L_3}^2-m_{E_3^c}^2) }{dt} 
&=&6\tilde{\alpha}_{2}M_{2}^{2} 
- \frac{6}{5}\tilde{\alpha}_{1}M_{1}^{2}
\nonumber \\
\frac{d(3m_{U_3^c}^2-2m_{H_{U}}^2  ) }{dt} 
&=& 16\tilde{\alpha}_{3}M_{3}^{2}
-6\tilde{\alpha}_{2}M_{2}^{2} 
+2\tilde{\alpha}_{1}M_{1}^{2}
\nonumber \\
\frac{d(3m_{D_3^c}^2+m_{E_3^c}^2-2m_{H_{D}}^2  ) }{dt} 
&=&16\tilde{\alpha}_{3}M_{3}^{2}
-6\tilde{\alpha}_{2}M_{2}^{2} 
+2\tilde{\alpha}_{1}M_{1}^{2}
\nonumber \\
\label{zeromodesMSSM}
\end{eqnarray}
Finally for completeness we include the MSSM RGEs for the $\mu$ and $B$ 
parameters:
\begin{eqnarray}
\frac{d\mu^2}{dt} &=&
(3\tilde{\alpha}_2 + \frac{3}{5} \tilde{\alpha}_1-3Y_t-3Y_b-Y_{\tau})\mu^2
\nonumber \\
\frac{dB}{dt} &=&
3\tilde{\alpha}_2M_2 + \frac{3}{5}\tilde{\alpha}_1M_1
-(3A_tY_t+3A_bY_b+A_{\tau}Y_{\tau})
\nonumber \\
\label{muBMSSM}
\end{eqnarray}

 In the present case we are interested in matching the high $\tan \beta$
MSSM with the strong unification predictions at the intermediate scale.
Thus the boundary conditions for the MSSM will involve {\em non-universal}
scalar masses, and unequal gauge coupling constants, 
although, to simplify the analysis, we shall
continue to use the good approximation 
of equal gaugino masses at the intermediate scale.
It is straightforward to find the
analytic solutions for the third family soft scalar mass RGEs
in the approximation $Y_t(t) \approx Y_b(t)$, which 
implies $A_t(t) \approx A_b(t)$, 
$m_{U_3^c}^2 (t)\approx m_{D_3^c}^2(t)$, 
and $m_{H_U}^2 (t)\approx m_{H_D}^2(t)$.
In our current notation 
$t=0$ now corresponds to the intermediate scale $M_I$ and $t=t$
corresponds to, say, the Z mass $M_Z$.
To be precise our notation is
\begin{eqnarray}
\alpha_3(M_{I}) & \equiv & \alpha_{3}(0)\nonumber \\
M_3(M_{I})  & \equiv & M_3(0)   \nonumber \\
m_i^2(M_{I})& \equiv & m_i^2(0)   \nonumber \\
\label{newnotation}
\end{eqnarray}

In this notation,
the solutions for the individual soft scalar masses may be expressed
in terms of the $\bar{m}_{U_3}^2(t)$ as follows:
\begin{eqnarray}
m_{Q_3}^2 (t)& = & 
\frac{1}{7}
\left[ 2\bar{m}_{U_3}^2(t)+5m_{Q_3}^2(0)-2m_{U_3^c}^2(0)-2m_{H_{U}}^2(0)\right]
\nonumber \\
& + & \frac{1}{7}M_3^2(0)\left[ 8\tilde{\alpha}_3(0)f_3
+\frac{9}{2}\tilde{\alpha}_2(0)f_2
-\frac{3}{2}\tilde{\alpha}_1(0)f_1\right]
\nonumber \\
m_{U_3^c}^2 (t)& = & 
\frac{1}{7}\left[
2\bar{m}_{U_3}^2(t)+5m_{U_3^c}^2(0)-2m_{Q_3}^2(0)-2m_{H_{U}}^2(0)\right]
\nonumber \\
& + & \frac{1}{7}M_3^2(0)\left[ 8\tilde{\alpha}_3(0)f_3
-6\tilde{\alpha}_2(0)f_2
+2\tilde{\alpha}_1(0)f_1\right]
\nonumber \\
m_{H_{U}}^2(t)& = & 
\frac{1}{7}\left[
3\bar{m}_{U_3}^2(t)+4m_{H_{U}}^2(0)-3m_{Q_3}^2(0)-3m_{U_3^c}^2(0)\right]
\nonumber \\
& + & \frac{1}{7}M_3^2(0)\left[ -16\tilde{\alpha}_3(0)f_3
+\frac{3}{2}\tilde{\alpha}_2(0)f_2
-\frac{1}{2}\tilde{\alpha}_1(0)f_1\right]
\label{msolnsMSSM}
\end{eqnarray}
where $f_i=f_i(t)$ are functions defined in \cite{ibanez}.

Most of the complication resides in the solution to
$\bar{m}_{U_3}^2(t)$ which is given below, in terms of the ratio
\begin{equation}
r_{QFP}\equiv \frac{Y_t(t)}{Y_{QFP}(t)}
\label{rQFP}
\end{equation}
and further functions defined in \cite{ibanez}
whose main property is that they do not depend on the Yukawa coupling:
\begin{eqnarray}
\bar{m}_{U_3}^2(t) & = & \bar{m}_{U_3}^2(0)(1-r_{QFP})
+M_3^2(0)\left[\frac{16}{3}\tilde{\alpha}_3(0)f_3
+3\tilde{\alpha}_2(0)f_2
+\frac{13}{15}\tilde{\alpha}_1(0)f_1\right]
\nonumber \\
&+& 3M_3^2(0)\left[{r_{QFP}}^2
\left( \frac{H_4^2}{3F^2}-\frac{2H_4H_2}{3F}+\frac{H_2^2}{3}\right)
+r_{QFP}\left(-G_1+\frac{G_2}{6F}-\frac{2H_2^2}{3}+\frac{2H_4H_2}{3F}
\right) \right]
\nonumber \\
&-&2\frac{H_3}{F}r_{QFP}(1-r_{QFP})M_3(0)A_t(0)
-r_{QFP}(1-r_{QFP})A_t(0)^2
\label{mbarsolnMSSM}
\end{eqnarray}

The remaining solutions are:
\begin{eqnarray}
Y_t(t) & = & \frac{Y_t(0)E}{1+7Y_t(0)F}
\nonumber \\
Y_{QFP}(t) & = & \frac{E}{7F}
\nonumber \\
1-r_{QFP} & = & \frac{1}{1+7Y_t(0)F}
\nonumber \\
A_t(t) & = & A_t(0)(1-r_{QFP})
+M_3(0)\left[H_2-r_{QFP}\left( \frac{tE}{F}-1\right) \right]
\nonumber \\
\mu^2(t)& = &\mu^2(0)(1+\beta_2t)^{\frac{3}{b_2}}
(1+\beta_1t)^{\frac{3}{5b_1}}(1-r_{QFP})^{\frac{6}{7}}
\nonumber \\
B(t)&=& B(0)+\frac{6}{7}r_{QFP}A_t(0)
+M_3(0)\left[\frac{6}{7}\left( \frac{tE}{F}-1\right)r_{QFP} -H_7\right]
\label{muBAsolnMSSM}
\end{eqnarray}

For the first two squark families (and for all three slepton families)
we may drop the Yukawa couplings to obtain the approximate solutions:
\begin{eqnarray}
m_{Q_i}^2 (t)& = & 
m_{Q_i}^2(0) + M_3^2(0)\left[ \frac{8}{3}\tilde{\alpha}_3(0)f_3
+\frac{3}{2}\tilde{\alpha}_2(0)f_2
+\frac{1}{30}\tilde{\alpha}_1(0)f_1\right]
\nonumber \\
m_{U_i^c}^2 (t)& = & m_{U_i^c}^2 (0)+ 
M_3^2(0)\left[ \frac{8}{3}\tilde{\alpha}_3(0)f_3
+\frac{8}{15}\tilde{\alpha}_1(0)f_1\right]
\nonumber \\
m_{D_i^c}^2 (t)& = & m_{D_i^c}^2 (0)+ 
M_3^2(0)\left[ \frac{8}{3}\tilde{\alpha}_3(0)f_3
+\frac{2}{15}\tilde{\alpha}_1(0)f_1\right]
\nonumber \\
m_{L_i}^2 (t)& = & 
m_{L_i}^2(0) + M_3^2(0)\left[ \frac{3}{2}\tilde{\alpha}_2(0)f_2
+\frac{3}{10}\tilde{\alpha}_1(0)f_1\right]
\nonumber \\
m_{E_i^c}^2 (t)& = & m_{E_i^c}^2 (0)+ 
M_3^2(0)\left[\frac{6}{5}\tilde{\alpha}_1(0)f_1\right]
\nonumber \\
\label{mlightMSSM}
\end{eqnarray}

The physical low-energy squark and slepton masses of all three families
also receive additional electroweak D-term contributions 
which for large $\tan \beta$ ($\cos 2\beta \approx -1$)
are as follows:
\begin{eqnarray}
m_{{U_L}_i}^2 (t)& = & m_{Q_i}^2(t)
+M_Z^2(-\frac{1}{2}+\frac{2}{3}\sin^2 \theta_W)
\nonumber \\
m_{{D_L}_i}^2 (t)& = & m_{Q_i}^2(t)
+M_Z^2(\frac{1}{2}-\frac{1}{3}\sin^2 \theta_W)
\nonumber \\
m_{{U_R}_i}^2 (t)& = & m_{U_i^c}^2 (t)
-M_Z^2(\frac{2}{3}\sin^2 \theta_W)
\nonumber \\
m_{{D_R}_i}^2 (t)& = & m_{D_i^c}^2 (t)
+M_Z^2(\frac{1}{3}\sin^2 \theta_W)
\nonumber \\
m_{{E_L}_i}^2 (t)& = & m_{L_i}^2(t)
+M_Z^2(\frac{1}{2}-\sin^2 \theta_W)
\nonumber \\
m_{{E_R}_i}^2 (t)& = & m_{E_i^c}^2 (t)+ M_Z^2(\sin^2 \theta_W)
\nonumber \\
m_{{N_L}_i}^2 (t)& = & m_{L_i}^2(t)-M_Z^2(\frac{1}{2})
\label{DMSSM}
\end{eqnarray}

We now have all the ingredients necessary
to calculate the low energy SUSY spectrum at the low energy scale $M_Z$
from the strong unification boundary conditions at the scale $M_I$
which is determined as a function of $n$ from Eq.\ref{MI}.
We assume gaugino dominance throughout so that the soft scalar masses at the
intermediate scale are family-independent and so the simple 
predictions in Eq.\ref{mTpredictions2} apply to each family separately:
\begin{eqnarray}
m_{Q_{i}}^{2}(0) &\approx &\frac{c_{Q}}{3}M_{1/2}^2  \nonumber \\
m_{U_{i}^{c}}^{2}(0) &\approx &\frac{c_{U}}{3}M_{1/2}^2  \nonumber \\
m_{D_{i}^{c}}^{2}(0) &\approx &\frac{c_{D}}{3}M_{1/2}^2  \nonumber \\
m_{L_{i}}^{2}(0) &\approx &\frac{c_{L}}{3}M_{1/2}^2  \nonumber \\
m_{E_{i}^{c}}^{2}(0) &\approx &\frac{c_{E}}{3}M_{1/2}^2  \nonumber \\
m_{H_{U}}^2(0) &\approx &\frac{c_{H_U}}{9}M_{1/2}^2  \nonumber \\
m_{H_{D}}^2(0) &\approx &\frac{c_{H_D}}{9}M_{1/2}^2
\label{mTpredictionsi}
\end{eqnarray}
The constants  $c_i$
have some model dependence, mainly through the parameter $n$,
but also through the nature of the Yukawa couplings.
We shall present results for the augmented model corresponding to the
coefficients in Table 3. Having fixed the high energy model,
the squark and slepton soft masses at the intermediate scale are thereby
also fixed, and so Eq.\ref{mTpredictionsi} with the coefficients in Table 3
may be used as input boundary conditions for the MSSM 
solutions in Eqs.\ref{msolnsMSSM} and \ref{mlightMSSM}.
The final ingredients are the fixed point prediction for the Yukawa 
couplings at the intermediate scale $Y_t(0)\approx Y_b(0)$,
which are given in terms of the QCD coupling at the intermediate
scale by Eq.\ref{FP's}. The QCD coupling at the intermediate
scale $\alpha_3(M_{I})$ is determined by
running up the LEP measured 
electroweak gauge couplings from $M_Z$ to $M_I$, and
$\alpha_3(M_{I})$ is then obtained from the fixed point prediction
in Eq.\ref{r}. The LEP measured QCD coupling is then predicted by 
running $\alpha_3(M_I)$ down to $M_Z$.

\begin{table}[tbp]
\hfil
\begin{tabular}{ccccccccc}
\hline
         & MSSM & MSSM & $n=40$ & $n=20$ & $n=20$ & $n=20$ & $n=20$ & $n=20$ 
\\ \hline
$M_{GUT}$   & $2\times 10^{16}$ & $2\times 10^{16}$ & $1\times 10^{16}$ &
$1\times 10^{16}$ & $7\times 10^{15}$ & $5\times 10^{15}$ & $3\times 10^{15}$ 
& $5\times 10^{15}$
\\ \hline
$\alpha_{GUT}$  & 1/24  & 1/24& 0.31   & 0.67 & 0.41  & 0.30  & 0.21 & 0.30  
\\ \hline
$\alpha_3(M_I)$   & - & - & 0.044 & 0.048 & 0.048 & 0.048 & 0.048 & 0.048
\\ \hline
$\alpha_3(M_Z)$  
& 0.121 & 0.121 & 0.114 & 0.114 & 0.114 & 0.114 & 0.114 & 0.114
\\ \hline
$M_{1/2}$    & 150   & 200 & 1000  & 1500 & 1000 & 1000 & 1000 & 750
\\ \hline
$M_3(M_I)$             & -   & -  & 143 & 108 & 118 & 162 & 228 & 121
\\ \hline
$\tilde{g}$          & 437   & 583 & 369   & 255 & 280  & 384  & 542 & 288
\\ \hline
$\tilde{\chi}_1^0$   & 59    & 80  & 67    & 59  & 64   & 88   & 124 & 65
\\ \hline
$\tilde{\chi}_2^0$   & 110   & 154 &  111   & 90 & 96   & 133  & 188 & 96
\\ \hline
$\tilde{\chi}_3^0$   & 261   & 347 &  289  & 457 & 331  & 361  & 420 & 273
\\ \hline
$\tilde{\chi}_4^0$   & 277   & 359 &  301  & 460 & 338  & 371  & 431 & 285
\\ \hline
$\tilde{\chi}_1^\pm$ & 110   & 154 &  110   & 90 & 96   & 132  & 189 & 95
\\ \hline
$\tilde{\chi}_2^\pm$ & 280   & 362 &  305  & 464 & 342  & 374  & 433 & 289
\\ \hline
$\tilde{E}_{L_i}$    & 119   & 153 &  209 & 397 & 273  & 282  & 301 & 214
\\ \hline
$\tilde{E}_{R_i}$    &  73   &  89 &   94   & 182 & 130  & 136  & 148 & 106
\\ \hline
$\tilde{N}_{L_i}$    &  88   &  130&   194  & 388 & 261  & 271  & 290 & 199
\\ \hline
$\tilde{U}_{L_i}$    &  398  & 532 &   377  & 455 & 361  & 431  & 550 & 321
\\ \hline
$\tilde{U}_{R_i}$    &  388  & 518 &   326  & 272 & 262  & 346  & 478 & 259
\\ \hline
$\tilde{D}_{L_i}$    &  406  & 538 &   385  & 462 & 370  & 438  & 555 & 331
\\ \hline
$\tilde{D}_{R_i}$    &  389  & 518 &   327  & 265 & 261  & 344  & 475 & 259
\\ \hline
$\tilde{t}_1$        &  212  & 326 &   200  & 224 & 183  & 240  & 337 & 155
\\ \hline
$\tilde{t}_2$        &  398  & 509 &   368 & 444 & 353  & 414  & 514 & 321
\\ \hline
$\tilde{b}_1$        &  245  & 353 &   192  & 155 &  138  & 220  & 335 & 135
\\ \hline
$\tilde{b}_2$        &  384  & 495 &   377  & 476 & 377  & 428  & 519 & 335
\\ \hline
$\mu(M_Z)$           & 250   & 338 & 278   & 449 & 320  & 353  & 413 & 261
\\ \hline
\end{tabular}
\hfil
\caption{\small Physical parameters and
SUSY masses (where applicable in GeV) based on the gaugino dominance 
assumption and assuming high $\tan \beta$. 
The first two columns of results correspond to the standard MSSM, assuming
gaugino dominance and high $\tan \beta$, for two different $M_{1/2}$ values.
The column labelled
$n=40$ corresponds to strong unification with an intermediate scale
of $M_I=3.7\times 10^{14}$ GeV and the columns labelled
$n=20$ correspond to strong unification with an intermediate scale
of $M_I=8\times 10^{12}$ GeV.
For strong unification we use the augmented model corresponding to the
$M_I$ boundary condition squark and slepton mass coefficients in Table 3.
The two input parameters are $M_{1/2}$ and
$\alpha_{GUT}$ which is controlled 
by a cut-off scale $M_{GUT}\equiv M_{NP}$.
All low energy masses are calculated at the scale $M_Z$,
including the parameter $\mu(M_Z)$ which is tuned so that the
correct tree-level electroweak symmetry breaking
condition $m_2^2\approx -\frac{M_Z^2}{2}$ is satisfied (see for example
\cite{wag})
where $m_2^2=m_{H_U}^2+\mu^2$. 
We set the high energy 
trilinear couplings equal to the high energy gaugino mass 
$A(M_{NP})=M_{1/2}$.}
\end{table}

The prediction of the gluino mass at the intermediate scale $M_3(M_I)$
depends on the high energy QCD coupling
$\alpha_3(M_{NP})$ which, being non-perturbative, is unknown.
For most of our predictions this does not matter
since the infra-red fixed point provides all the low energy predictions.
But the gaugino masses are not determined by the fixed point, and so their
low energy values are sensitive to the high energy (non-perturbative) physics.
We have already assumed for simplicity that the gaugino masses are unified
in the infra-red (down to $M_I$) but to relate $M_3(M_I)$
to $M_{1/2}\equiv M_3(M_{NP})$ requires techniques beyond perturbation theory.
In this paper we shall parametrise this relationship by the
value of $\alpha_3(M_{NP})$ which we shall regard as a free parameter,
and determine $M_3(M_I)$ from $M_{1/2}$ from the one-loop
result in Eq.\ref{wellknown}. Of course we do not trust perturbation 
theory up to $M_{NP}$, but this proceedure is sufficient to parametrise
the uncertainties in $M_3(M_I)$ for a given $M_{1/2}$.
Thus the two basic input parameters in our approach may be regarded as
$M_{1/2}\equiv M_3(M_{NP})$
and $\alpha_{GUT}\equiv \alpha_3(M_{NP})$.
In practice we control the value of $\alpha_{GUT}$ by introducing
a high energy cut-off scale which we denote as
$M_{GUT}\equiv M_{NP}$ where the evolution of the
couplings is arbitrarily stopped. The gauge couplings are not of course
unified at $M_{GUT}\equiv M_{NP}$ since they remain in their
fixed point ratios at this scale, and this scale has no physical
significance; it is simply a cut-off scale for our high energy theory,
which may be higher or lower depending on how much faith you have
in perturbation theory.
\footnote{In the two or three loop RGE analysis the value of 
$M_{GUT}\equiv M_{NP}$ will be predicted to be somewhat
different from the typical values
in Table 5 \cite{strong1, kolda}. Similarly we would expect a more refined
numerical analysis of the soft scalar masses to yield a somewhat modified
estimate of the SUSY spectrum than in Table 5.
However the basic features of the spectrum are 
captured by the present one loop analysis.}

In Table 5 we present the SUSY spectrum for both the MSSM and
strong unification, in both cases assuming large $\tan \beta$
and making the assumption of gaugino dominance. 
We give the QCD coupling and the
gluino mass at $M_{GUT}\equiv M_{NP}$,
at the intermediate scale $M_I$, and at low energies $M_Z$.
It is observed that strong unification
predicts a smaller value of $\alpha_3(M_Z)$ than the MSSM.
The MSSM spectrum always involves light sleptons $\tilde{E}_{R_i}$
whose mass cannot be increased by further increasing
$M_{1/2}$ since this would lead to its becoming the lightest
supersymmetric particle (LSP). 
By contrast the strong unification model with $n=40$ has a slightly
heavier slepton mass but lighter stop and sbottom masses
around 200 GeV, while the other strong unification models have a
comfortably heavy slepton mass accompanied by reasonably light stops and
sbottoms. The third family top, bottom and tau masses have already been
discussed elsewhere using more accurate two-loop RGEs \cite{strong3}.
We only remark that we would expect the top mass to be somewhat smaller
in the augmented model than in the original model
due to the fixed point ratios in Eq.\ref{FP's} being smaller than
the values in Eq.\ref{FPs}, and this will reduce the quoted top quark masses
in Table 2 of ref.\cite{strong3}, bringing them closer
to the experimentally observed value.

For a given value of $M_{1/2}$ and strong gauge coupling 
$\alpha_{GUT}$ the SUSY masses are increased slightly by taking smaller
values of $n$, as seen by comparing the results for $n=40$ to those
in the $\alpha_{GUT}=0.3$ column corresponding to $n=20$,
but the effect is not great. 
A larger effect is to change the value of $\alpha_{GUT}$ as is done for 
the $n=20$ model by systematically reducing the value of $M_{GUT}$
from $10^{16}$ GeV to $3\times 10^{15}$ GeV, which results
in $\alpha_{GUT}$ being reduced from 0.67 to 0.21 over this range.
Since $\alpha_3(M_I)$ is unchanged, smaller values of $\alpha_{GUT}$
result in larger values of low energy gluino mass and consequently a heavier 
SUSY spectrum, with the effect being felt most directly for the
lighter gauginos (which are predominantly wino and bino due to the
relatively large $\mu$ parameter) but also for the
squarks and sleptons whose masses receive gaugino corrections.
The SUSY spectra which correspond to having the 
lightest stops and sbottoms for a given chargino mass are achieved
by having the smallest values of $M_{1/2}$ for a given
value of $\alpha_{GUT}$, as in the case of the last column
with $M_{1/2}=750$ GeV. The uncertainties in the low energy gluino mass
for a given $M_{1/2}$ should not mask the relations in Eq.\ref{mTpredictionsi}
which are the main predictions of our scenario. The range
of spectra in Table 5 reflect the uncertainty in the gluino mass
and give the general features of the spectrum expected in this model
for a range of parameters.

Finally we remark that we have not included any analysis of the
spectrum of Higgs masses in this model since their treatment
lies outside of our simple approximations here. For example one
must examine the differences between $Y_t$ and $Y_b$, and also 
one must seriously consider the question of radiative corrections
to the Higgs potential which are very important in determining the
Higgs masses. We merely remark that with high $\tan \beta$ we would
expect the Higgs masses to be heavier than for low $\tan \beta$,
which could help to explain why the lightest Higgs 
has yet to be observed at LEP2.

\section{D-term flavour violation}

In this section
we make some remarks about flavour violation in the model \cite{IR} arising
from the presence of the gauged $U(1)_{X}$ family symmetry. Here we face two
problems: firstly such gauged family symmetries are expected to give rise to
flavour changing effects via the D-terms; secondly this symmetry is
anomalous and must be broken close to the string scale. Since the expansion
parameter for fermion masses is given by this scale divided by $M_{I}$ this
apparently requires $M_{I}$ to also very close to the string scale which in
turn would seem to require very large values of $n$. (For example $%
M_{I}=10^{13}$ GeV for $n=20,$ higher values of $M_{I}$ require larger
values of $n$). We shall comment on each of these problems in turn.

First let us explain the problem with the D-term. Consider a model with the
$U(1)_{X}$ anomalous gauge symmetry of reference \cite{IR}. 
There are two MSSM singlet
fields $\theta ^{(1)}$ and $\bar{\theta}^{(-1)}$ which couple to MSSM
non-singlet fields (quarks,leptons,Higgs and heavy vector reps). For example
the D-term including the contribution of the quark doublets is: 
\begin{equation}
\frac{1}{2}g_{X}^{2}D^{2}=\frac{1}{2}%
g_{X}^{2}(-4|Q_{1}^{(-4)}|^{2}+|Q_{2}^{(1)}|^{2}+|\theta ^{(1)}|^{2}-|\bar{%
\theta}^{(-1)}|^{2}+\xi )^{2}
\end{equation}
where the anomalous term is 
\begin{equation}
\xi =\frac{g_{X}^{2}}{192\pi ^{2}}(TrX)M_{P}^{2}
\end{equation}
Typically $\sqrt{\xi }\sim 0.1M_{string}$ which, as we shall see, will
set the scale for $X$ symmetry breaking.

Assuming that all $Q$ VEVs are zero, the potential we need to minimise will
only involve $\theta^{(1)}$ and $\bar{\theta}^{(-1)}$ fields: 
\begin{equation}
V_0= \frac{1}{2}g_X^2(|\theta^{(1)} |^2 - |\bar{\theta}^{(-1)} |^2 +\xi )^2
+m^2 |\theta^{(1)} |^2 + \bar{m}^2 |\bar{\theta}^{(-1)} |^2
\end{equation}
where at $M_{string}$, 
\begin{equation}
m^2=\bar{m}^2=m_{3/2}^2
\end{equation}

This potential is minimised by a $\bar{\theta}^{(-1)}$ VEV with a zero $%
\theta ^{(1)}$ VEV: 
\[
<\theta ^{(1)}>=0,\ \ \ \ <|\bar{\theta}^{(-1)}|^{2}>=\xi -\frac{\bar{m}^{2}%
}{g_{X}^{2}} 
\]
and $X$ symmetry will be broken at a scale $\sqrt{\xi }$. Replacing the $%
\theta ^{(1)}$ and $\bar{\theta}^{(-1)}$ fields by their VEVs we generate
the D-term contribution to the squark mass terms: 
\begin{equation}
\frac{1}{2}g_{X}^{2}D^{2}=\frac{1}{2}%
g_{X}^{2}(-4|Q_{1}^{(-4)}|^{2}+|Q_{2}^{(1)}|^{2}+\frac{\bar{m}^{2}}{g_{X}^{2}%
})^{2}
\end{equation}
which gives soft squark mass contributions of order $\bar{m}^{2}$ (the $%
g_{X} $ coupling cancels). These mass contributions are dangerous because
they depend on the family X charge so are family dependent. In the limit
that $m^{2}=\bar{m}^{2}=0$ the symmetry breaking will still take place,
since it is driven by the anomaly, but the $D$ term will be exactly zero and
no contributions to squark masses will occur. This is important because it
means that if there were some reason why $\bar{m}^{2}$ were small at the
symmetry breaking scale, then the worrysome D-term contributions would also
be small. In particular if more than one field acquires a $U(1)_{X}$
breaking vev the magnitude of the residual D-term will be governed by the 
{\it lightest} field. In this case there is an obvious candidate for an
anomously light field playing this role since moduli fields $\phi $ may also
carry $X$ charge and may contribute to the D term. Such moduli fields do not
have bare soft mass terms. Radiative corrections involving gauge couplings
will induce moduli masses as in Eq(\ref{zero}) but due to the extremely
large value of the $U(1)_{X}$ beta function\cite{LR,LR2} they will be very
small.

To illustrate how this mechanism works we include a pair of moduli fields $%
\phi _{+}$ and $\phi _{-}$ with charges $X=\pm 1$, respectively. Assuming
all squark VEVs and the $\theta $ VEV are zero, and ignoring the moduli
mass, the tree level potential is: 
\begin{equation}
V_{0}=\frac{g_{X}^{2}}{2}(\xi -\bar{\theta}^{2}-\phi _{-}^{2}+\phi
_{+}^{2})^{2}+\bar{m}^{2}\bar{\theta}^{2}
\end{equation}
where we have just taken the real components of $\bar{\theta},\phi $, and
added a soft mass for $\bar{\theta}$, taking the moduli fields $\phi^\pm $ to
have zero mass for simplicity. Radiative symmetry breaking
corresponds to the mass squared $\bar{m}^{2}$ being driven negative, at some
MS-bar scale $\mu _{0}$, resulting in a $\bar{\theta}$ VEV. If we work at
the scale $\mu _{0}$ then the tree-level potential will receive one-loop
corrections of the form: 
\begin{equation}
\Delta V_{1}=\left[ \frac{d\bar{m}^{2}}{d\ln \mu ^{2}}\right] _{\mu _{0}}%
\bar{\theta}^{2}\ln (\bar{\theta}^{2}/\mu _{0}^{2})
\end{equation}
Minimising the potential at the scale $\mu _{0}$, and including the one-loop
corrections, we find that 
\begin{eqnarray}
<\bar{\theta}^{2}> & \approx & \mu _{0}^{2} \nonumber \\
<\phi _{-}^{2}>-<\phi _{+}^{2}> & = & \xi -<\bar{\theta}^{2}>
\label{VEVs}.
\end{eqnarray}
Thus both the $\bar{\theta}$ 
field and the moduli fields are expected to develop VEVs.
Moreover, inserting these VEVs back into the potential, we see that the
D-term contribution to the squark masses is zero. If we now include small
moduli masses then we find that the D-term contribution to squark masses is
controlled by the small moduli masses. As a result we can simultaneously
avoid the appearance of large family asymmetric D-terms and separate the
scale of the $\bar{\theta}$ VEV (which is determined by $\mu _{0}$) from the
scale of the symmetry breaking of the anomalous symmetry.

Of course there remains the question of the $\theta \,\,$VEV. In the above
we assumed for simplicity
that the ${\theta}\,\ $VEV was smaller than $\mu
_{0}.$ For large $n$ this need not be the case and then minimising the
effective potential, including the $\theta $ field, shows that $\bar{\theta}%
\approx \theta \approx \mu _{0}$ where now $\mu _{0}$ is the scale at which
the sum of the $\theta $ and $\bar{\theta}$ masses squared becomes negative.
If the $\bar{\theta}\,\ $ VEV is larger than $\mu _{0\ }$ there may
still be a $\theta $ VEV but it will be generated at the scale at which the 
$\theta $ field mass squared is driven negative.

We conclude this section by summarising the various mass scales 
we have introduced.
The basic scale is the high energy string scale, which is of order the
scale $M_{NP}$. The intermediate scale $M_I$, which may be much smaller, 
is generated by a radiative symmetry breaking mechanism involving additional
singlets $\Phi$ 
which are responsible for generating the masses of the additional
vector representations \cite{strong1}. (As already noted the masses of the 
vector representations may somewhat exceed the VEVs of such singlets.)
A similar
radiative mechanism is also responsible for the $\theta$, $\bar{\theta}$,
$\phi^{\pm}$ VEVs just discussed. Just as the intermediate scale may be
much smaller than the string scale, so it is possible that the 
$\theta$ and $\bar{\theta}$ VEVs are much smaller than the 
$\phi^{\pm}$ VEVs. Indeed the model of fermion masses which motivates the
present analysis requires that the parameter 
$\epsilon \approx <\theta >/M_{I} \approx <\bar{\theta}>/M_{I}\approx 0.2$.
Thus for small values of $n$, where the intermediate scale is 
necessarily small, we rely on the radiative mechanism to generate
large hierarchies of VEVs of the kind we desire.

\newpage

\section{Summary and Conclusions}

In this paper we have examined in some detail the implications for the soft
supersymmetry breaking mass terms of the infra-red fixed points of the
renormalisation group equations. The model examined has the property that
the Yukawa couplings were driven to family independent fixed points. This
suggested that the soft supersymmetry breaking masses might similarly be
driven to family independent fixed points providing a new solution to the
flavour changing problem. However this analysis has shown that
the individual soft masses are not in fact driven to fixed points,
but instead only certain combinations of soft masses are driven
to fixed points. The main reason for this is that
the renormalisation group equations have several anomalous
dimensions which vanish. As a result the values of the soft terms at lower
energies depend on some of the initial values, 
so that the flavour problem is not dynamically solved but requires the initial
conditions should be approximately family independent. The appearance of
these troublesome zero anomalous dimensions 
appears generic because the right-hand side
of the renormalisation group equations involve common combinations of the
soft terms and not the soft terms individually. Thus one can eliminate these
terms through suitable combination of the 
renormalisation group equations leaving 
combinations of the masses renormalisation group invariant.
\footnote{A model with a larger gauge group in which the quarks, leptons
and Higgses are treated symmetrically such as trinification, may
lead to soft mass fixed points.} 

A general solution for the soft masses was obtained in terms of a 
limited number of initial conditions. Although the individual 
soft masses are not driven to fixed points,
the combinations of masses appearing in Eq.\ref{X} are 
driven to the fixed points:
\begin{eqnarray}
m_{Q_{i}}^{2}(M_I)+m_{U_{j}^{c}}^{2}(M_I)+m_{H_{U_{ij}}}^{2}(M_I) 
&=& M_3^2(M_I)
\nonumber \\
m_{Q_{i}}^{2}(M_I)+m_{D_{j}^{c}}^{2}(M_I)+m_{H_{D_{ij}}}^{2}(M_I)
&=& M_3^2(M_I)
\nonumber \\
m_{L_{i}}^{2}(M_I)+m_{E_{j}^{c}}^{2}(M_I)+m_{H_{D_{ij}}}^{2}(M_I)
&=& M_3^2(M_I)
\label{X1}
\end{eqnarray}
where we have assumed the relations $m_1=m_2=1$ corresponding to
gaugino unification at the fixed point, and used the trilinear
fixed point relations.
Moreover, since the gluino mass at the 
intermediate scale is driven to be small compared
to its value at the initial scale, these combinations are relatively small.
Since the masses appearing in this equation 
evolve at different rates these constraints 
have the important effect that some soft 
mass squared are driven negative at the fixed points.

In order to address the flavour problem we 
explored a particularly attractive scheme 
in which the gaugino masses dominate at the unification scale. 
In this case the fixed point structure proves to be very predictive. 
The assumption of gaugino dominance at the unification 
scale implies that initially soft masses 
and hence family differences between 
soft masses are small. In running down to the intermediate
scale the family sums of soft masses is
increased to values of order the initial high energy gluino mass $M_{1/2}$,
i.e.
$m_T^2(M_I)=cM_{1/2}^2$ where $c$ are simple numerical 
coefficients which we tabulated in Tables 2 and 3.
On the other hand the differences of soft masses at the intermediate
scale are not enhanced. 
Thus gaugino mass dominance leads to a strong 
relative suppression of the family dependence of the
scalar masses and an associated strong 
suppression of flavour changing neutral currents.
Since gaugino mass dominance implies all families 
have approximately equal soft masses, the fixed points for the
sums of
soft masses in Eq.\ref{X1} apply to each family separately
\begin{eqnarray}
m_{Q_{i}}^{2}(M_I)+m_{U_{i}^{c}}^{2}(M_I)+m_{H_{U}}^{2}(M_I) 
&=& M_3^2(M_I)
\nonumber \\
m_{Q_{i}}^{2}(M_I)+m_{D_{i}^{c}}^{2}(M_I)+m_{H_{D}}^{2}(M_I)
&=& M_3^2(M_I)
\nonumber \\
m_{L_{i}}^{2}(M_I)+m_{E_{i}^{c}}^{2}(M_I)+m_{H_{D}}^{2}(M_I)
&=& M_3^2(M_I)
\label{Xi}
\end{eqnarray}
Here we have replaced the Higgs by the MSSM Higgs doublets $H_U$ and $H_D$
since all Higgs doublets will have equal soft masses.
This is a strong fixed point prediction in the gaugino dominance
limit. Furthermore, since $M_3^2(M_I)$ is driven small,  the
combinations of masses on the left hand side are approximately zero 
and hence we are able to predict the
individual soft scalar masses in terms of the initial gaugino mass 
as in Eq.\ref{mTpredictionsi}. 
In the simple model originally considered, corresponding to 
the coefficients $c_i$ in Table 2,
some of the mass squared
terms of this squarks and sleptons were driven negative at the fixed point
signalling a breakdown of colour and/or electroweak symmetry at high energies.
However we have argued that the extra Higgs couplings
necessarily present in a realistic model of the type we have considered
will provide the cure to this problem. To illustrate the effect
we discussed a model with three additional quark and lepton
families coupling to the Higgs sector, and shown that the mass squareds
of the squarks and sleptons remain positive, as in Table 3.

Given the large negative Higgs mass squareds at the intermediate scale
one might worry about the question of 
so-called ``unbounded from below'' constraints which have recently been
revisited \cite{Abel2}. The worry can most easily be revealed by looking
at the MSSM just below the intermediate scale where the high energy
fixed point analysis predicts the combinations of soft masses in 
Eq.\ref{Xi} to be approximately zero.
Since we have arranged for the squark and sleptons to have positive
mass squared, one concludes that combinations like
$m_{L_3}^2+m_{H_U}^2$ are negative near the intermediate scale
which immediately implies that the flat direction corresponding
to $L_3Q_3D^c_3$ and $H_{U}L_3$ develops a dangerous minimum \cite{Abel2}.
This conclusion seems to be an unavoidable consequence of the fixed
point structure of the theory. However, as emphasised by the authors 
themselves \cite{Abel2}, the tunneling time from the desired MSSM vacuum
to the flat direction vacuum always exceeds the age of the universe.
Thus whether or not we would be sitting in the wrong vacuum
depends on cosmological details of the history of the universe.
For example many theories of inflation tend to lift flat directions
by amounts of order the Hubble constant. This could provide
a useful mechanism to roll the universe into the correct MSSM vacuum.

The combination of the fixed point structure and
gaugino dominance leads to a low energy spectrum quite unlike 
that predicted by gauge dominance in the MSSM, since the
gluino mass at the intermediate scale is expected to be small compared
to the soft masses at this scale. Detailed low energy spectra based on
gaugino dominance are presented in Table 5, both for strong unification
and for the MSSM for comparison (both for large $\tan \beta$.) 
The spectra
corresponding to values of gaugino masses consistent with current
limits and yet having light stops and sbottoms, correspond to taking 
the smaller values of $M_{1/2}$, as for example in the last column
of Table 5 with $M_{1/2}=750$ GeV. Whereas in the gauge dominated MSSM 
the right-handed sleptons are rather light, here they are comfortably heavy.
Although the soft scalar masses at the intermediate scale are
precisely predicted in terms of $M_{1/2}$ by Eq.\ref{mTpredictionsi},
the relation between the soft gaugino masses at the intermediate scale 
and $M_{1/2}$ is less clear, and this uncertainty is reflected in the range
of spectra in Table 5. Generally, however, we would expect a
``scalar dominated'' SUSY spectrum characterised by large soft scalar
masses and large $\mu$ parameter, with smaller soft gaugino masses.

Finally we discussed the additional source of flavour changing
arising from the D-term of the gauged family symmetry. Such gauged family
symmetries are commonly introduced in models of fermion masses, and so
this problem is generic. 
We pointed out that such D-term problems 
are less severe than hitherto thought if anomalously
light fields such as moduli have family gauge quantum numbers.

We conclude that in strong unification 
the combination of the fixed point structure
with the assumption of gaugino 
dominance leads to a predictive and successful scheme where
flavour changing neutral currents are heavily suppressed,
squark and slepton soft masses at the intermediate scale are
predicted in terms of $M_{1/2}$, and the low energy spectrum is
phenomenologically viable.

We are grateful to the Aspen Center for Physics where some of this work was
carried out.

\end{document}